\documentclass[12pt,english,longbibliography,nofootinbib,superscriptaddress,12pt,sort&compress,showkeys]{revtex4}
\usepackage[T1]{fontenc}
\usepackage[utf8]{inputenc}
\usepackage{xcolor}
\usepackage{amsmath}
\usepackage{amssymb}
\usepackage{graphicx}
\usepackage{esint}
\PassOptionsToPackage{normalem}{ulem}
\usepackage{ulem}

\makeatletter

\providecommand{\tabularnewline}{\\}
\providecolor{lyxadded}{rgb}{0,0,1}
\providecolor{lyxdeleted}{rgb}{1,0,0}

\DeclareRobustCommand{\lyxsout}[1]{\ifx\\#1\else\sout{#1}\fi}

\@ifundefined{textcolor}{}
{%
 \definecolor{BLACK}{gray}{0}
 \definecolor{WHITE}{gray}{1}
 \definecolor{RED}{rgb}{1,0,0}
 \definecolor{GREEN}{rgb}{0,1,0}
 \definecolor{BLUE}{rgb}{0,0,1}
 \definecolor{CYAN}{cmyk}{1,0,0,0}
 \definecolor{MAGENTA}{cmyk}{0,1,0,0}
 \definecolor{YELLOW}{cmyk}{0,0,1,0}
}

\usepackage{indentfirst}
\usepackage{amsfonts}
\usepackage{ae,aecompl}
\usepackage{sidecap}
\usepackage[section]{placeins}
\usepackage{epsf}
\makeatother

\usepackage{babel}
\date{\today}

\usepackage{chngcntr}

\makeatother

\usepackage{babel}
\begin{document}
\title{Development of a general-purpose machine-learning interatomic potential
for aluminum by the physically-informed neural network method}
\author{G.P. Purja Pun}
\address{Department of Physics and Astronomy, MSN 3F3, George Mason University,
Fairfax, Virginia 22030, USA}
\author{V. Yamakov}
\address{National Institute of Aerospace, Hampton, Virginia 23666, USA}
\author{J. Hickman}
\address{Materials Science and Engineering Division, National Institute of
Standards and Technology, Gaithersburg, Maryland 20899-8910, USA}
\author{E. H. Glaessgen}
\address{NASA Langley Research Center, Hampton, Virginia 23681, USA}
\author{Y. Mishin}
\address{Department of Physics and Astronomy, MSN 3F3, George Mason University,
Fairfax, Virginia 22030, USA}
\begin{abstract}
Interatomic potentials constitute the key component of large-scale
atomistic simulations of materials. The recently proposed physically-informed
neural network (PINN) method combines a high-dimensional regression
implemented by an artificial neural network with a physics-based bond-order
interatomic potential applicable to both metals and nonmetals. In
this paper, we present a modified version of the PINN method that
accelerates the potential training process and further improves the
transferability of PINN potentials to unknown atomic environments.
As an application, a modified PINN potential for Al has been developed
by training on a large database of electronic structure calculations.
The potential reproduces the reference first-principles energies within
2.6 meV per atom and accurately predicts a wide spectrum of physical
properties of Al. Such properties include, but are not limited to,
lattice dynamics, thermal expansion, energies of point and extended
defects, the melting temperature, the structure and dynamic properties
of liquid Al, the surface tensions of the liquid surface and the solid-liquid
interface, and the nucleation and growth of a grain boundary crack.
Computational efficiency of PINN potentials is also discussed.
\end{abstract}
\maketitle

\section{Introduction\label{sec:Introduction}}

Large-scale molecular dynamics (MD) and Monte Carlo (MC) simulations
constitute an essential component of the multiscale approach in materials
modeling and computational design. The most critical ingredient of
such simulations is the classical interatomic potentials, whose role
is to make computationally fast predictions of the system energy and
atomic forces. It is not an exaggeration to say that the results of
atomistic simulations are as accurate and reliable as the utilized
interatomic potentials. Several forms of interatomic potentials have
been developed for different classes of materials. Some of the most
popular types of potentials include the embedded-atom method (EAM)
potentials \citep{Daw84,Daw83,Mishin.HMM}, the modified embedded-atom
method (MEAM) potentials \citep{Baskes87}, the angular-dependent
potentials \citep{Mishin05a}, the charge-optimized many-body potentials
\citep{Liang:2012aa}, the reactive bond-order potentials \citep{Brenner90,Brenner00,Stuart:2000aa},
and the reactive force fields \citep{van-Duin:2001aa} -- to name
a few. During the past decade, a new class of machine-learning (ML)
potentials has emerged, which is based on a radically different philosophy
than the traditional potentials.

The traditional interatomic potentials partition the total energy
$E$ into energies $E_{i}$ assigned to individual atoms $i$: $E=\sum_{i}E_{i}$.
Each atomic energy $E_{i}$ is expressed as a function of atomic positions
$(\mathbf{r}_{i1},\mathbf{r}_{i2},...,\mathbf{r}_{in})$ in the vicinity
of atom $i$. This function depends on a small number of fitting parameters
$\mathbf{p}=(p_{1},...,p_{m})$, which are optimized on a database
composed of experimental data and a relatively small set of energies
and/or forces obtained by electronic structure calculations. Once
optimized, the potential parameters are fixed once and for all and
used for all atomic environments that might be encountered during
the subsequent MD and/or MC simulations. Traditional potentials are
computationally fast and scale linearly with the number of atoms.
As such, they provide access to systems containing millions of atoms
and enable MD simulations for tens or even hundreds of nanoseconds.
Because they are based on a small number of parameters, the accuracy
of traditional potentials is generally not very high. However, the
functional form of traditional potentials is motivated by physical
understanding of the interatomic bonding in the material in question.
As a result, the potentials often demonstrate reasonable transferability
to atomic configurations that were not included in the fitting database.
Although the energies and forces predicted outside the fitting domain
may not be very accurate, they retain some degree of physical sense.
Another feature of the traditional potentials is that they are typically
general-purpose type. Once released to the community, a potential
is used not only for the purpose for which it was intended but for
almost any type of simulations that the user might wish to perform.

The emerging class of ML potentials takes a different approach. The
physics of interatomic bonding is not considered. The local environment
of an atom is mapped directly onto the potential energy surface (PES)
using one of the high-dimensional nonlinear regression methods, such
as the Gaussian process regression \citep{Payne.HMM,Bartok:2010aa,Bartok:2013aa,Li:2015aa,Glielmo:2017aa,Bartok_2018,Deringer:2018aa},
the kernel ridge regression \citep{Botu:2015bb,Botu:2015aa,Mueller:2016aa},
or an artificial neural network (NN) \citep{Behler07,Bholoa:2007aa,Behler:2008aa,Sanville08,Eshet2010,Handley:2010aa,Behler:2011aa,Behler:2011ab,Sosso2012,Behler:2015aa,Behler:2016aa,Schutt:148aa,Imbalzano:2018aa}.
Other types of ML potentials include the spectral neighbor analysis
(SNAP) \citep{Thompson:2015aa,Chen:2017ab,Li:2018aa} and moment tensor
(MTP) \citep{Shapeev:2016aa} potentials. In most cases, the total
energy is again partitioned into atomic contributions. However, instead
of position vectors $(\mathbf{r}_{i1},\mathbf{r}_{i2},...,\mathbf{r}_{in})$
of neighboring atoms, a set of local structural parameters $(G_{i}^{1},G_{i}^{2},...,G_{i}^{K})$
is introduced, which encodes the local environment of the atom and
is invariant under rotations and translations of the coordinate axes.
The approach based on local descriptors was pioneered by Behler and
Parrinello \citep{Behler07} (who called $G_{i}^{j}$ the symmetry
parameters \citep{Behler07,Behler:2011ab}) in the context of NN potentials.
Since the size $K$ of the feature vector $(G_{i}^{1},G_{i}^{2},...,G_{i}^{K})$
is fixed, a single NN can be trained for all atoms of the system.
The NN (or any other regression model) contains a large number of
parameters ($\approx10^{3}$), which are trained on a large database
of first-principles energies and/or forces (typically, for $\approx10^{3}$
to $\approx10^{4}$ supercells). A high accuracy of fitting can be
achieved, usually on the meV per atom level. The required reference
data can be generated by high-throughput density functional theory
(DFT) calculations.

The method has a wide scope of applications since the regression and
its training do not depend of the nature of chemical bonding in the
material. However, the high accuracy and flexibility come at a price:
the ML potentials suffer from poor transferability to atomic configurations
lying outside the training domain. Since the structure-energy mapping
is not guided by any physics or chemistry, the regression only ensures
accurate numerical interpolation between the DFT points. Extrapolation
outside the domain of known environments is purely mathematical and
thus cannot be expected to make physically meaningful predictions.
The lack of physics-based transferability presents a challenge to
the development of general-purpose type ML potentials.

The recently proposed physically-informed neural network (PINN) model
\citep{PINN-1} aims to improve the transferability of ML potentials
by integrating a NN regression with a physics-based interatomic potential.
Whereas the parameters of a traditional potential are permanently
fixed, the PINN model predicts the best set of potential parameters
for every atomic environment that may be encountered during simulations.
To achieve this, the local structural descriptors $(G_{i}^{1},G_{i}^{2},...,G_{i}^{K})$
are fed into a pre-trained NN, which outputs an optimized set of potential
parameters $\mathbf{p}_{i}=(p_{i1},...,p_{im})$ for the given atom
$i$. These parameters are then used to compute the atomic energy
$E_{i}$ with the potential. The atomic energies are summed up to
obtain the total energy of the system. Like the mathematical ML potentials
mentioned above, the PINN model predicts the PES of the system, from
which analytical forces acting on the atoms can be obtained by differentiation.
In other words, the PINN model relies on a physics-based interatomic
potential, but the potential parameters are adjusted on the fly by
a NN according to local environments of atoms. Improved transferability
to new environments is expected because extrapolation is now guided
by physical insights embedded in the interatomic potential rather
than a purely numerical algorithm.

In the previous paper \citep{PINN-1}, a PINN potential for aluminum
(Al) was constructed as a proof of principle. The goal of the present
paper is to report on further developments of the PINN method and
to construct and test a new, significantly improved version of the
general-purpose PINN Al potential.

In Section \ref{sec:Methodology}, we briefly review the PINN model
and describe its modifications in both the core formalism and the
training/validation procedures. In Section \ref{sec:PINN_Al}, we
present the new Al potential with a superior quality over the previous
version \citep{PINN-1}. We test the potential for a wide spectrum
of physical properties, such as lattice dynamics, thermal expansion,
defect structures and energies in the face-centered-cubic (FCC) Al,
and equations of state of several alternate crystalline phases of
Al. Next, we apply the potential to compute several properties that
require extrapolation to diverse environments and can only be obtained
by large-scale simulations (Section \ref{sec:Tests}). The applications
include structural and dynamic properties of liquid Al, the melting
temperature of Al, as well as the surface tensions of the liquid surface
and the solid-liquid interface. Another application involving almost
half a million atoms is the growth of a crack on a planar grain boundary.
While performing these tests, we evaluate the computational efficiency
of the PINN simulations -- an important topic that we discuss in
Section \ref{sec:Computational-efficiency}. In Section \ref{sec:Conclusions},
we summarize this work.

\section{Methodology\label{sec:Methodology}}

\subsection{The bond-order potential}

The key ingredients of the PINN model are a physics-based interatomic
potential, local structural parameters (descriptors), and an artificial
NN connecting the descriptors and the potential parameters. We will
start by discussing the interatomic potential.

\begin{figure}
(a) \includegraphics[width=0.28\textwidth]{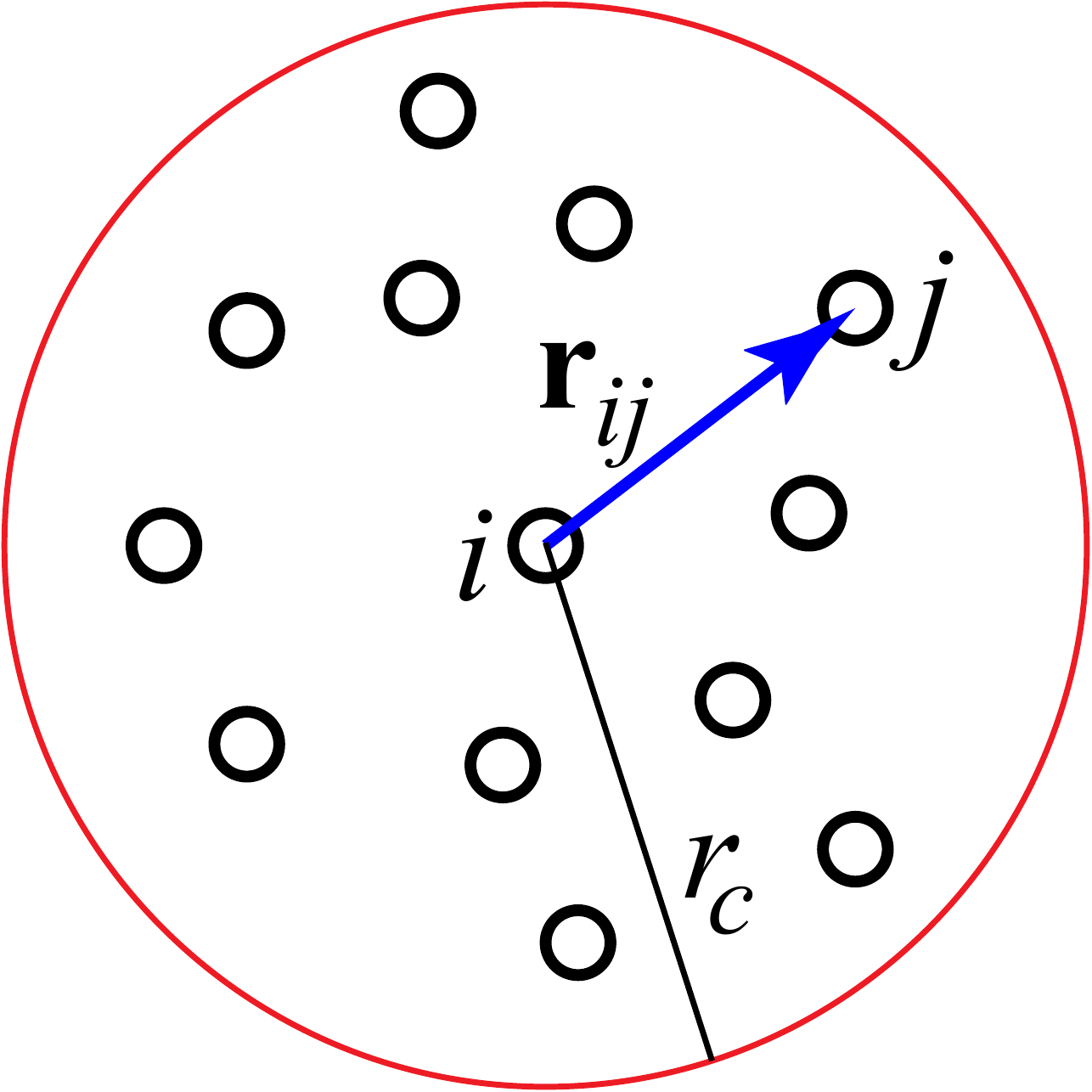}

\bigskip{}

(b) \includegraphics[width=0.37\textwidth]{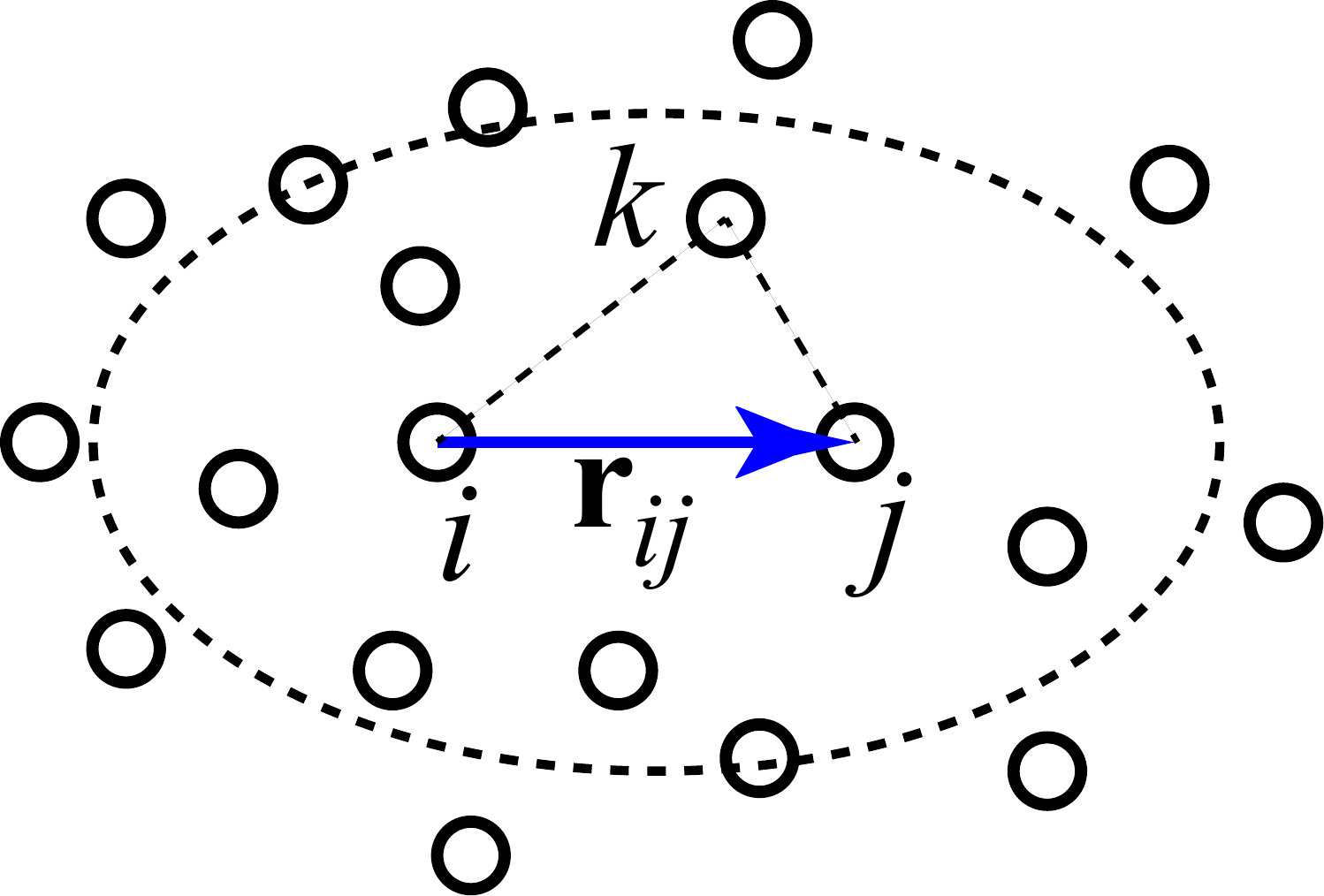}

\bigskip{}

(c) \includegraphics[width=0.37\textwidth]{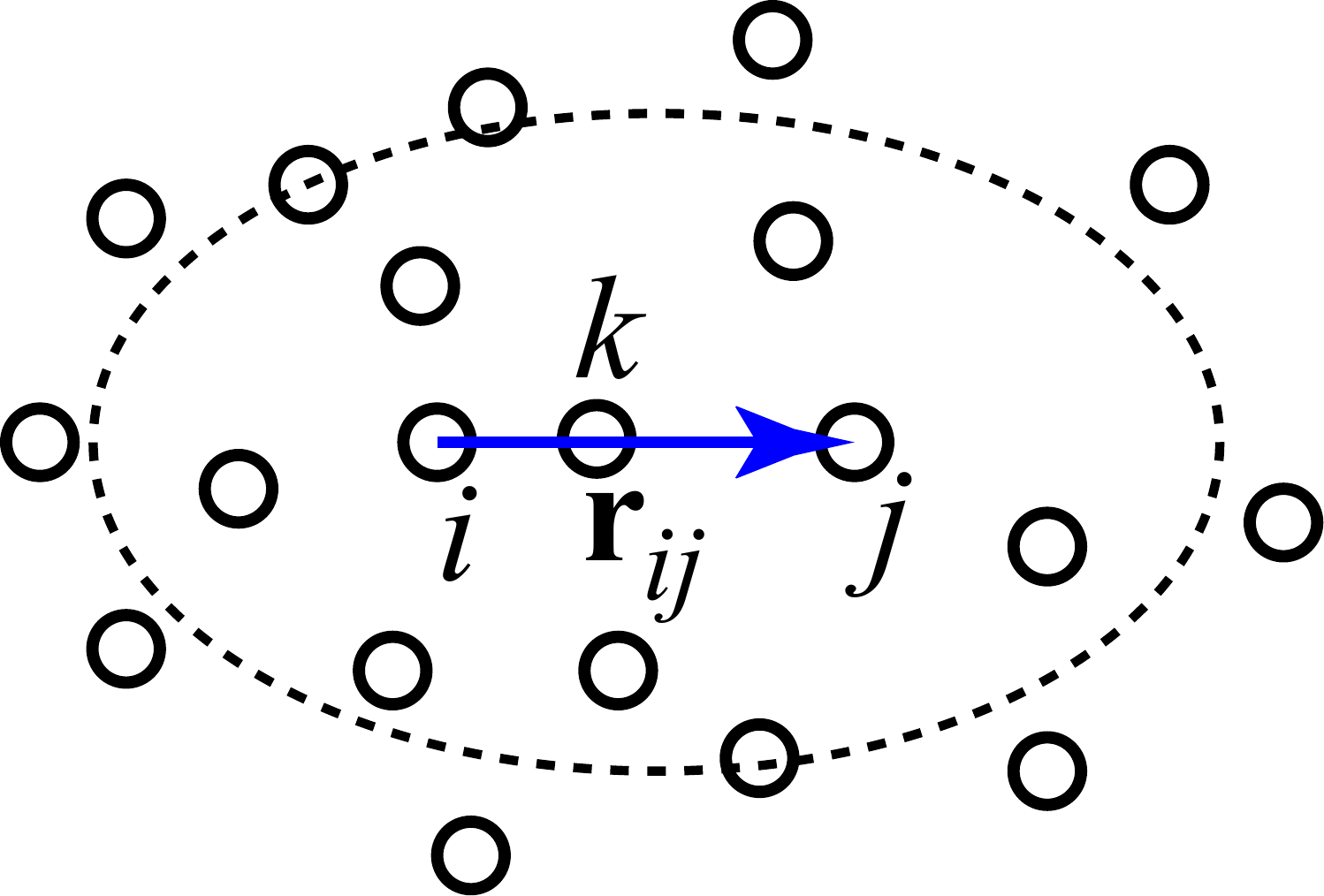}

\caption{(a) Neighbors $j$ and an atom $i$ within the cutoff sphere of radius
$r_{c}$. (b) Atomic bond $i$-$j$ is partially screened by surrounding
atoms $k$. The surfaces of constant screening factor are ellipsoids
whose poles coincide with the atomic positions $i$ and $j$. (c)
If an atom $k$ is located on the bond $i$-$j$, then the screening
factor is close to zero and the bond is broken. \label{fig:BOP}}
\end{figure}

As in \citep{PINN-1}, we choose an analytical bond-order potential
(BOP) \citep{Oloriegbe_PhD:2008aa,Gillespie:2007vl,Drautz07a} capable
of describing chemical bonding in both covalent and metallic materials.
For a single-component system, the total energy $E=\sum_{i}E_{i}$
is the sum of the atomic energies
\begin{equation}
E_{i}=\dfrac{1}{2}\sum_{j\ne i}\left[e^{A_{i}-\alpha_{i}r_{ij}}-S_{ij}b_{ij}e^{B_{i}-\beta_{i}r_{ij}}\right]f_{c}(r_{ij},d,r_{c})+E_{i}^{(p)},\label{eq:BOP1}
\end{equation}
where $A_{i}$, $B_{i}$, $\alpha_{i}$, $\beta_{i}$, $d$ and $r_{c}$
are parameters. The summation runs over neighbors $j$ of atom $i$.
$r_{ij}$ is the distance between the two atoms. The interactions
are truncated at a distance $r_{c}$ using the cutoff function
\begin{equation}
f_{c}(r,r_{c},d)=\begin{cases}
\dfrac{(r-r_{c})^{4}}{d^{4}+(r-r_{c})^{4}}\enskip & r\leq r_{c}\\
0,\enskip & r\geq r_{c},
\end{cases}\label{eq:BOP2}
\end{equation}
where the parameter $d$ controls the truncation smoothness. The cutoff
sphere encompasses several coordination shells and typically contains
a few dozen atoms (Fig.~\ref{fig:BOP}a).

In Eq.(\ref{eq:BOP1}), the first term in the square brackets describes
the repulsion between neighboring atoms at short separations, whereas
the second term describes the chemical bonding. The coefficient 
\begin{equation}
b_{ij}=(1+z_{ij})^{-1/2},\label{eq:BOP3}
\end{equation}
captures the bond-order effect. Here $z_{ij}$ represents the number
of bonds $ik$ formed by the atom $i$ (not counting the bond $i$-$j$,
which is included by adding the unity). The bonds are counted with
weights that depend on the angle $\theta_{ijk}$ between the bonds
$ij$ and $ik$:
\begin{equation}
z_{ij}=\sum_{k\ne i,j}a_{i}S_{ik}\left(\cos\theta_{ijk}-h_{i}\right)^{2}f_{c}(r_{ik},d,r_{c}),\label{eq:BOP4}
\end{equation}
where $a_{i}$ and $h_{i}$ are parameters. The angular dependence
is introduced to capture the directional character of covalent bonds.
According to Eq.(\ref{eq:BOP3}), atoms surrounded by a larger number
of neighbors have a lower energy per bond (the bond order effect).

All chemical bonds are screened by the screening factor $S_{ij}$
defined by
\begin{equation}
S_{ij}=\prod_{k\ne i,j}S_{ijk},\label{eq:BOP5}
\end{equation}
where the partial screening factors $S_{ijk}$ represent the contributions
of individual atoms $k$ to the screening of the bond $i$-$j$. The
partial screening factors have the form
\begin{equation}
S_{ijk}=1-f_{c}(r_{ik}+r_{jk}-r_{ij},d,r_{c})e^{-\lambda_{i}(r_{ik}+r_{jk}-r_{ij})},\label{eq:BOP6}
\end{equation}
where $\lambda_{i}$ is the screening parameter (inverse of the screening
length). Eq.(\ref{eq:BOP6}) shows that $S_{ijk}$ has a constant
value on a spheroid whose poles coincide with atoms $i$ and $j$
(Fig.~\ref{fig:BOP}b). The cutoff spheroid is defined by the condition
$r_{ik}+r_{jk}-r_{ij}=r_{c}$ and encompasses all atoms $k$ that
contribute to the screening. For atoms $k$ located outside the cutoff
spheroid $S_{ijk}=1$ (no contribution to the screening), while for
atoms inside cutoff spheroid $S_{ijk}<1$. The closer the atom $k$
to the bond $i$-$j$, the smaller is $S_{ijk}$ and the larger is
its contribution to the screening. If one of the atoms $k$ is located
on the bond $i$-$j$ (Fig.~\ref{fig:BOP}c), then $r_{ik}+r_{jk}-r_{ij}=0$
and $S_{ijk}=1-f_{c}(0,d,r_{c})\ll1$. Thus, the bond $i$-$j$ is
almost completely screened and can be considered as broken. The deviation
from complete screening ($S_{ijk}=0$) is controlled by the parameter
$d$ in Eq.(\ref{eq:BOP2}) and avoids division by zero in the analytical
differentiation of the potential.

It should be noted that the major semi-axis of the cutoff spheroid
has the length $1.5r_{c}$, i.e., is larger than the radius $r_{c}$
of the cutoff sphere. Thus, some atoms lying outside the cutoff sphere
can still affect the atomic energy $E_{i}$ indirectly through the
screening effect.

The last term in Eq.(\ref{eq:BOP1}) is an on-site energy given by
\begin{equation}
E_{i}^{(p)}=-\sigma_{i}\left({\displaystyle \sum_{j\neq i}S_{ij}b_{ij}}f_{c}(r_{ij})\right)^{1/2},\label{eq:BOP7}
\end{equation}
$\sigma_{i}$ being a parameter. For covalent materials, $E_{i}^{(p)}$
is added to account for the promotion energy required to change the
electronic structure of free atoms when they form covalent bonds.
For metallic materials, $E_{i}^{(p)}$ represents the energy of embedding
the atom into the local electron density. Indeed, $E_{i}^{(p)}$ can
be recast in the form
\begin{equation}
F(\bar{\rho}_{i})=-\sigma_{i}\left(\bar{\rho}_{i}\right)^{1/2},\label{eq:BOP8}
\end{equation}
where
\begin{equation}
\bar{\rho}_{i}=\sum_{j\neq i}S_{ij}b_{ij}f_{c}(r_{ij})\label{eq:BOP9}
\end{equation}
has the meaning of the host electron density on atom $i$. This term
is similar to the embedding energy $F(\overline{\rho})$ appearing
in the EAM method widely used for metallic systems.

Thus, the BOP potential underlying the PINN model reflects the nature
of chemical bonding in both covalent and metallic materials. As such,
it can be employed for the modeling of mixed-bonding materials and
multi-phase systems containing metal-nonmetal interfaces.

The BOP potential depends on 10 parameters. Eight of them, namely,
$A_{i}$, $B_{i}$, $\alpha_{i}$, $\beta_{i}$, $a_{i}$, $h_{i}$,
$\sigma_{i}$ and $\lambda_{i}$ are adjusted according to the local
atomic environments.\footnote{Note that the definitions of $a_{i}$ and $\lambda_{i}$ are different
from those in the original PINN formulation \citep{PINN-1}.} The cutoff parameters $r_{c}$ and $d$ are treated as global: once
adjusted, they remain the same for all atoms.

\subsection{The local structural parameters}

The local environment of an atom $i$ is encoded in the rotationally-invariant
structural parameters
\begin{equation}
g_{i}^{(l)}(r_{0},\sigma)=\sum_{j\neq i,k\neq i}P_{l}\left(\cos\theta_{ijk}\right)f(r_{ij},r_{0},\sigma)f(r_{ik},r_{0},\sigma),\enskip\enskip l=0,1,2,...,l_{max},\label{eq:5-1-1}
\end{equation}
where $P_{l}(x)$ are Legendre polynomials of order $l$. The radial
function is the Gaussian 
\begin{equation}
f(r,r_{0},\sigma)=\dfrac{1}{r_{0}}e^{-(r-r_{0})^{2}/\sigma^{2}}f_{c}(r,1.5r_{c},d)\label{eq:5-1}
\end{equation}
of width $\sigma$ centered at point $r_{0}$. Note that the truncation
radius for this function is $1.5r_{c}$ to capture the positions of
atoms $j$ and $k$ lying outside the cutoff sphere of the potential
but affecting the atomic energy through the screening.

A set of Gaussian parameters $\left\{ r_{0}^{(n)},\sigma^{(n)}\right\} $,
$n=1,2,...,n_{max}$, is selected and the coefficients $\sinh^{-1}\left[g_{i}^{(l)}(r_{0}^{(n)},\sigma^{(n)})\right]$
are arranged in an array $\mathbf{G}_{i}=(G_{i}^{1},G_{i}^{2},...,G_{i}^{K})$
of the fixed length $K=l_{max}n_{max}$. This array serves as the
feature vector representing the environment.

\subsection{The neural network and its training}

In the initial PINN formulation \citep{PINN-1}, the NN regression
mapped the vector of local structural parameters $\mathbf{G}_{i}(\mathbf{r}_{i1},\mathbf{r}_{i2},...,\mathbf{r}_{in})$
onto a set of BOP parameters $\mathbf{p}_{i}$: $\mathbf{G}_{i}\underset{NN}{\rightarrow}\mathbf{p}_{i}$.
These parameters were then used to compute the atomic energy $E_{i}(\mathbf{r}_{i1},\mathbf{r}_{i2},...,\mathbf{r}_{in},\mathbf{p}_{i})$.
Mathematically, the energy calculation can be expressed by the composite
function
\begin{equation}
E_{i}=E_{i}\left(\mathbf{r}_{i1},\mathbf{r}_{i2},...,\mathbf{r}_{in},\underbrace{\mathbf{p}_{i}\left(\mathbf{G}_{i}(\mathbf{r}_{i1},\mathbf{r}_{i2},...,\mathbf{r}_{in})\right)}_{NN}\right).\label{eq:NN1}
\end{equation}

\begin{figure}
\noindent \begin{centering}
\includegraphics[width=0.85\textwidth]{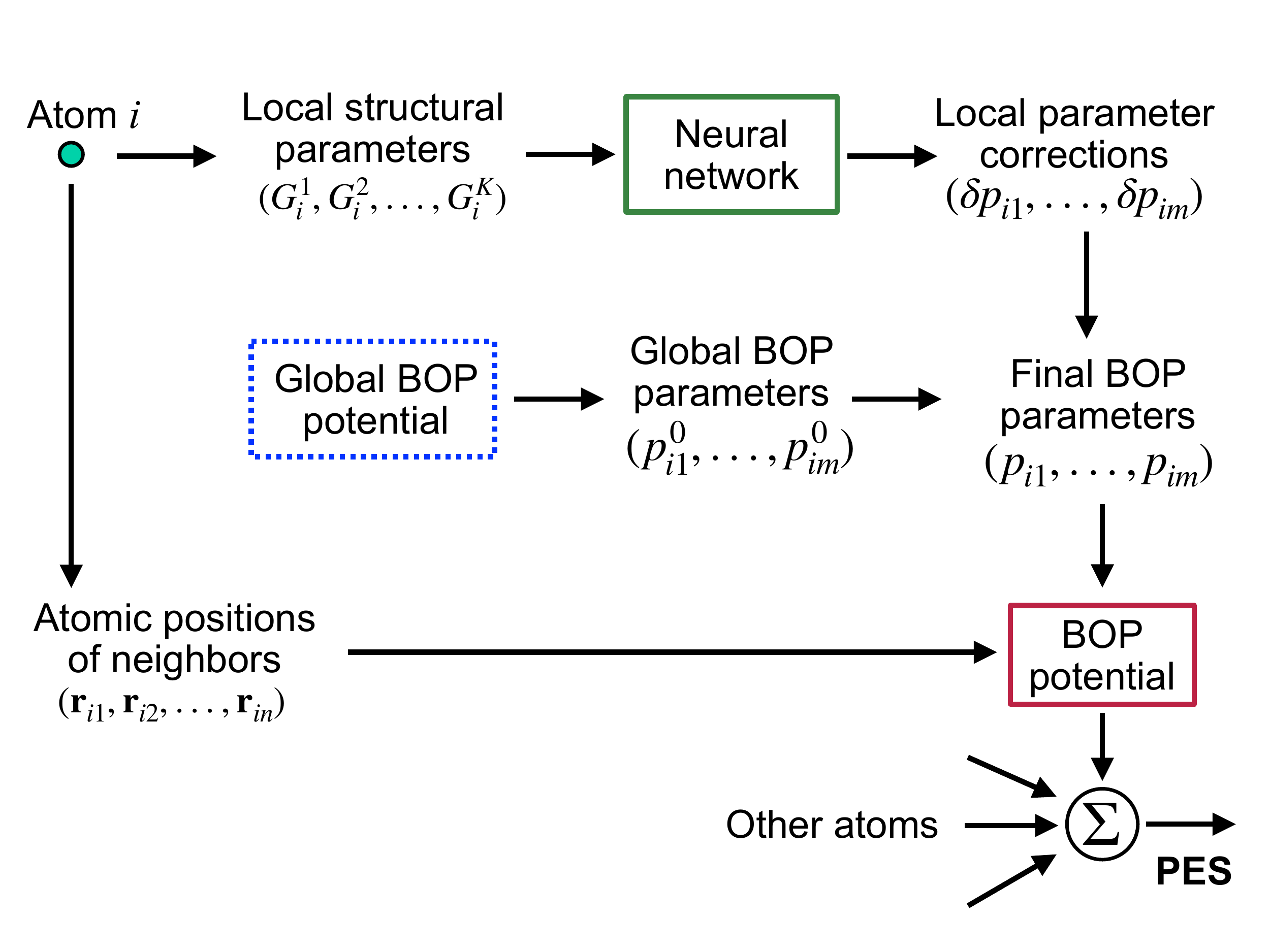}
\par\end{centering}
\caption{Flowchart of the modified PINN method. The notations are explained
in the text.\label{fig:Flowchart-PINN}}
\end{figure}

In the modified version of PINN presented here, the starting point
is a global BOP potential whose parameters have been trained on the
entire DFT database. Let the optimized set of BOP parameters obtained
be denoted $\mathbf{p}^{0}$. Since this set of parameters is small,
the root-mean-square error (RMSE) of fitting is not expected to be
low. Rather, it is usually on the order of $10^{2}$ meV per atom.
Next, a pre-trained NN adds to $\mathbf{p}^{0}$ a set of local ``perturbations''
$\delta\mathbf{p}_{i}=(\delta p_{i1},...,\delta p_{im})$ such that
the final parameter set $\mathbf{p}_{i}=\mathbf{p}^{0}+\delta\mathbf{p}_{i}$
predicts the energy $E_{i}$ with much better accuracy. Mathematically,
the modified PINN formula is 
\begin{equation}
E_{i}=E_{i}\left(\mathbf{r}_{i1},\mathbf{r}_{i2},...,\mathbf{r}_{in},\mathbf{p}^{0}+\underbrace{\delta\mathbf{p}_{i}\left(\mathbf{G}_{i}(\mathbf{r}_{i1},\mathbf{r}_{i2},...,\mathbf{r}_{in})\right)}_{NN}\right).\label{eq:NN2}
\end{equation}
The diagram in Fig.~\ref{fig:Flowchart-PINN} explains the flow of
information in the method. Note that the role of the atomic coordinates
$(\mathbf{r}_{i1},\mathbf{r}_{i2},...,\mathbf{r}_{in})$ of the neighboring
atoms is twofold: they are arguments of the BOP potential, and they
are also used to compute the local structural parameters which, in
turn, predict the local corrections $\delta\mathbf{p}_{i}$ to the
global BOP parameters $\mathbf{p}^{0}$ after passing through the
NN.

In the proposed scheme, the energy predictions are largely guided
by the global BOP potential, whose role is to provide a smooth and
physically meaningful extrapolation outside the training domain. The
magnitudes of the weights and biases of the NN can be controlled to
keep the local corrections $\delta\mathbf{p}_{i}$ as small as possible.
This approach is designed to improve the transferability of the PINN
potential while keeping a high level of accuracy. Tests also show
that the modified scheme improves the stability and the speed of convergence
during the NN training.

Although the NN is allowed to have any architecture and size, we find
that a simple feedforward network with 2 to 3 hidden layers is sufficient
for achieving the desired accuracy of training. The input and output
layers contain $K$ (number of descriptors) and 8 (number of BOP parameters)
nodes, respectively. The loss function has the form
\begin{eqnarray}
\mathcal{E} & = & \dfrac{1}{N}\sum_{s}\left(\dfrac{E^{s}-E_{\textrm{DFT}}^{s}}{N_{s}}\right)^{2}+\tau_{1}\dfrac{1}{N_{p}}\left(\sum_{\epsilon\kappa}\left|w_{\epsilon\kappa}\right|^{2}+\sum_{\nu}\left|b_{\nu}\right|^{2}\right)\nonumber \\
 & + & \tau_{2}\dfrac{1}{N_{a}m}\sum_{s}\sum_{i_{s}=1}^{N_{s}}\sum_{n=1}^{m}\left|p_{i_{s}n}-\overline{p}_{i_{s}n}\right|^{2}+\tau_{3}\dfrac{1}{N_{a}m}\sum_{s}\sum_{i_{s}=1}^{N_{s}}\sum_{n=1}^{m}\left|p_{i_{s}n}\right|^{2}\label{eq:NN3}
\end{eqnarray}
Here $E^{s}$ is the total energy of supercell $s$ predicted by the
potential, $E_{\textrm{DFT}}^{s}$ is the DFT energy of this supercell,
$N_{s}$ is the number of atoms in the supercell, $N$ is the total
number of supercells in the database, $N_{a}$ is the total number
of atoms in all supercells, $w_{\epsilon\kappa}$ and $b_{\nu}$ are
the weights and biases of the NN, $N_{p}$ is the total number of
NN parameters, and $\tau_{1}$, $\tau_{2}$ and $\tau_{3}$ are adjustable
coefficients. The first term in the right-hand side is our definition
of the mean-square error of fitting. The remaining terms are added
for regularization purposes. The second term ensures that the network
parameters remain reasonably small for smooth interpolation. The third
term controls variations of the BOP parameters relative to their values
$\overline{p}_{i_{s}n}$ averaged over the training database. The
last term is optional and was added to prevent the BOP parameters
from growing beyond physically reasonable limits.

Because of the complex structure of the PINN potential and the loss
function, application of the standard NN training methods such as
backpropagation is impractical. Instead, we implement the Davidon-Fletcher-Powell
algorithm of unconstrained optimization \citep{NRC2} in the high-dimensional
space of the NN parameters ($N_{p}\gg1$). This algorithm requires
the knowledge of partial derivatives of $\mathcal{E}$ with respect
to the NN parameters, which were derived analytically by multi-step
application of the chain rule. The global BOP potential is optimized
by the same algorithm. The loss function has many local minima, hence
the training has to be repeated multiple times starting from different
initial conditions. Due to the large size of the optimization problem,
the training process relies on massive parallel computations as will
be discussed later.

\section{Development of the PINN potential for Al \label{sec:PINN_Al}}

\subsection{The potential training and validation}

The PINN Al potential developed here was trained, validated and tested
on the same DFT database as was used in the original version \citep{PINN-1}.
The training and validation database, which we denote $\mathcal{D}$,
was composed of DFT energies of 36,490 supercells. These supercells
represented seven crystal structures of Al under isotropic and uniaxial
tensions and compressions, surfaces with different crystallographic
orientations, five symmetrical tilt grain boundaries, unrelaxed intrinsic
stacking fault, a vacancy, and several isolated clusters containing
from 2 to 79 atoms. Some of the supercells were static (0 K temperature),
but most of them were snapshots of DFT MD simulations at different
atomic volumes and temperatures. A database $\mathcal{F}\subset\mathcal{D}$
composed of 3,164 supercells (108,052 atoms) randomly selected from
$\mathcal{D}$ was created for training purposes. The structures included
in the training database $\mathcal{F}$ are described in detail in
the Supplementary File accompanying this paper. In addition, 10 more
datasets $\mathcal{V}_{i}$, each containing 495 supercells (19,540
atoms), were randomly selected from $\mathcal{D}$ for validation
purposes. These validation datasets lay outside the training database
($\mathcal{V}_{i}\subset\mathcal{D}\setminus\mathcal{F}$) and did
not intersect with each other ($\mathcal{V}_{i}\cap\mathcal{V}_{j}=\textrm{Ø}$).
They were used to control overfitting during the training process.
Yet another DFT database $\mathcal{T}$ composed of 26,425 supercells
(2,376,388 atoms) was used for testing the potential as will be discussed
later. This database was composed of structures different from those
in the training and validation database ($\mathcal{T}\cap\mathcal{D}=\textrm{Ø}$).
More detailed information about the databases and the DFT calculations
can be found in the Supplementary File and in \citep{PINN-1}.

A number of different descriptors $\mathbf{G}_{i}$, network architectures
(including variations in the number of neurons in the hidden layers)
and regularization parameters were tested. In each case, the NN weights
and biases were initialized by random numbers in the interval {[}-0.1,0.1{]}.
The optimizer had to be restarted about 10 times with different initial
conditions to avoid early trapping in a local minimum. While it was
almost always possible to train the model to the same RMSE of about
3 meV per atom, the predicted physical properties varied significantly
from one potential to another. Even with all hyper-parameters fixed,
the potentials trained to the same RMSE starting from different random
sets of weights and biases predicted slightly different sets of physical
properties. The final version of the potential selected for this paper
was obtained with $\tau_{1}=10^{-4}$, $\tau_{2}=0$, $\tau_{3}=0.02$,
$r_{c}=6$\,Å, $d=1.5$\,Å and $\sigma=1$\,Å. The feature vector
has the size of $K=40$ corresponding to five Legendre polynomials
of orders $l=0,1,2,4$ and 6 with 8 Gaussian positions at $r_{0}=\{2.0,2.5,3.0,3.5,4.0,5.0,6.0,7.0\}$\,Å
(see Eqs.(\ref{eq:5-1-1}) and (\ref{eq:5-1}) for notation). The
NN architecture is $40\times16\times16\times8$ with a total of $N_{p}=1064$
fitting parameters. The two hidden layers contain 16 nodes each, and
the output layer contains $m=8$ nodes (number of local BOP parameters).
The RMSE of training is 2.60 meV per atom. During the training, the
RMSE's of the validation datasets $\mathcal{V}_{i}$ were recorded
to make sure that the potential is not subject to overfitting or selection
bias. The validation errors continually decreased during the training
process, as shown on the convergence plot in the Supplementary File
(increase in the validation error would signal overfitting). For the
final potential, the RMSE of validation averaged over the 10 validation
sets was 3.94 meV per atom.

Figure \ref{fig:compVsDFT} shows that the potential predictions are
in excellent agreement with the DFT energies uniform across the 7
eV per atom wide energy range. The error distribution is centered
at zero (no bias) and has an approximately Gaussian shape. For comparison,
Fig.~\ref{fig:BOP-DFT} shows the energies predicted by the global
BOP potential plotted against the DFT energies. The BOP potential
generally follows the DFT trend but is less accurate than the PINN
potential and displays significant deviations for some of the high-energy
structures. The plot demonstrates the drastic improvement in accuracy
achieved by the local adjustments of the BOP parameters implemented
in the PINN potential.

\begin{figure}[h!]
\noindent \begin{centering}
(a) \includegraphics[scale=0.32]{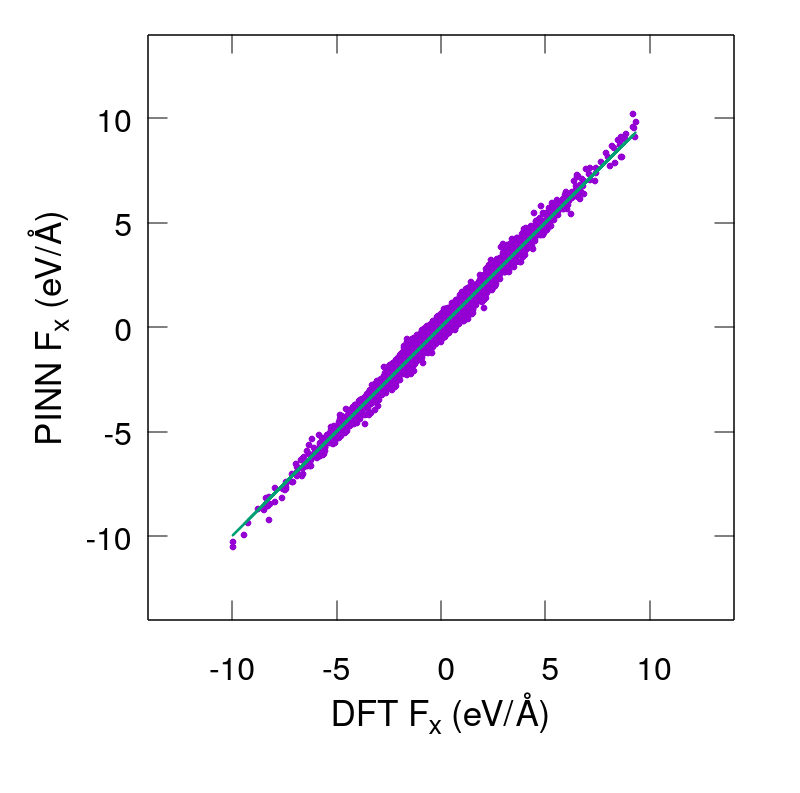}
\par\end{centering}
\noindent \begin{centering}
(b) \includegraphics[scale=0.32]{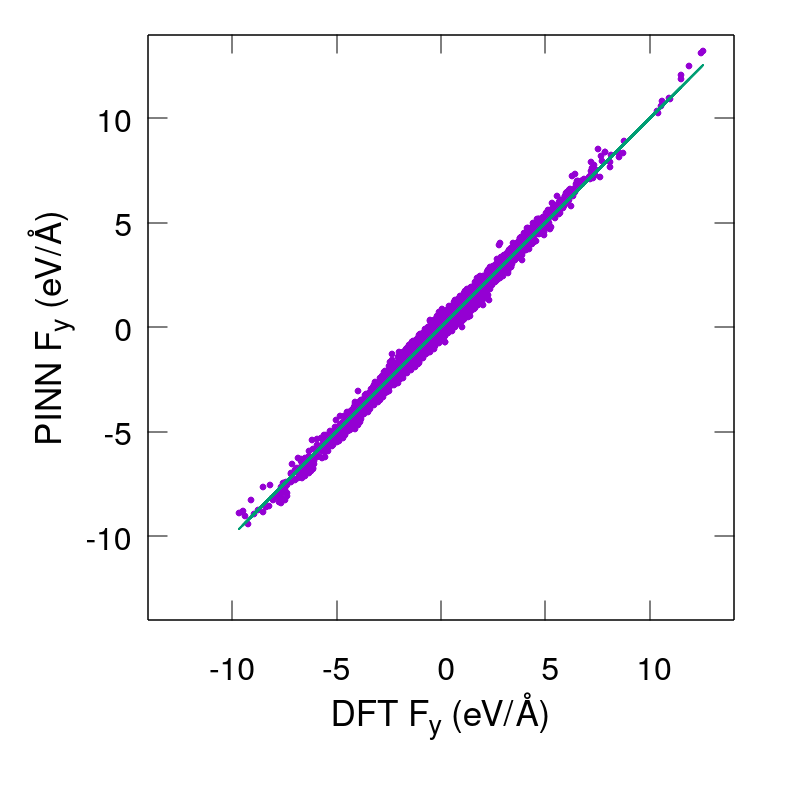}
\par\end{centering}
\noindent \centering{}(c) \includegraphics[scale=0.32]{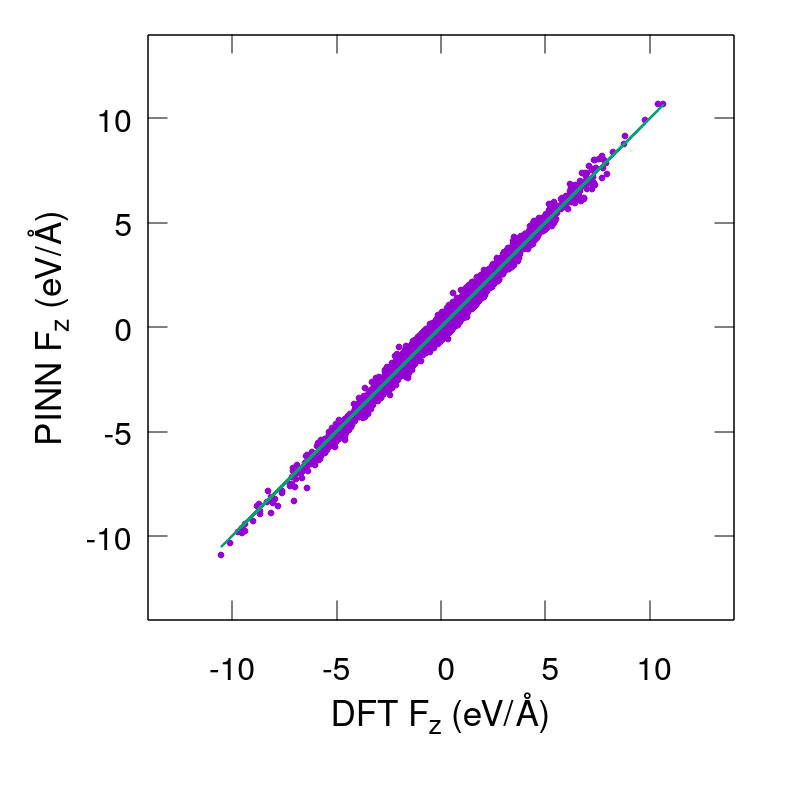}
\caption{Components of the atomic forces predicted by the PINN potential in
comparison with DFT calculations for the training database. The straight
lines represent the perfect fit. The RMS deviation is 0.11 eV\,Å$^{-1}$.
DFT forces were not used during the potential training and validation.
\label{Fig:Force-training}}
\end{figure}

DFT forces were not used during the training and validation, but they
were checked against the potential predictions once the final version
was selected. The potential forces display an unbiased scatter relative
to the DFT forces with the RMSE of about 0.1 eV\,Å$^{-1}$ (Fig.~\ref{Fig:Force-training}).
Forces are sometimes included in the training of ML potentials. This
option will be explored with other PINN potentials in the future.
In the present case, we chose to examine the forces after the training
to demonstrate that the potential was not overfitted (overfitting
would manifest itself in a small error in energies and a large error
in forces).

\subsection{Tests of basic properties}

Properties of Al predicted by the PINN potential were computed with
the ParaGrandMC (PGMC) code developed at the National Aeronautics
and Space Administration (NASA) \citep{ParaGrandMC,Purja-Pun:2015aa,Yamakov:2015aa}.
The code implements massively parallel MD and MC simulations in a
variety of statistical ensembles. It works with several types of interatomic
potentials, including the modified PINN potential described in this
work. The atomic forces and the stress tensor are computed from analytical
expressions. Further details related to this code will be discussed
in Section \ref{sec:Computational-efficiency}. The atomic structures
appearing in the paper were analyzed and visualized using the Open
Visualization Tool (OVITO) visualization tool \citep{Stukowski2010a}.

\begin{figure}[htpb]
\noindent \begin{centering}
\includegraphics[scale=0.45]{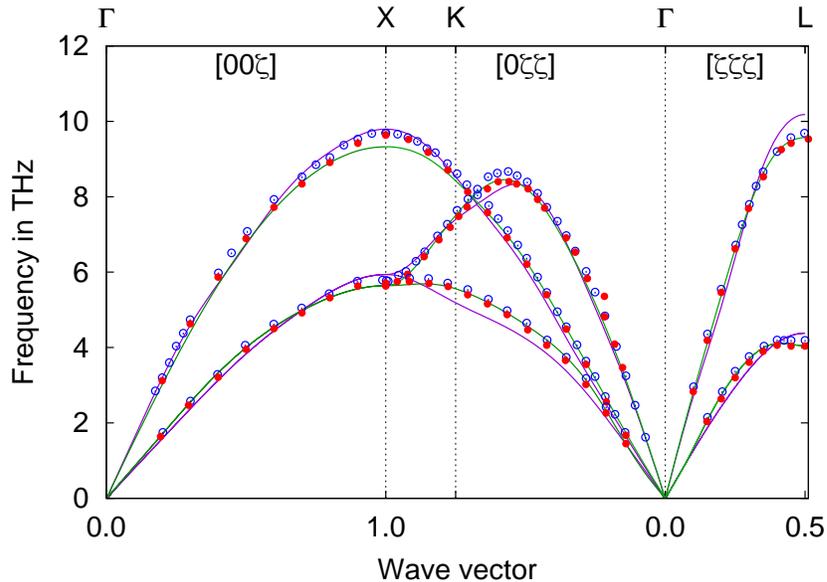}
\par\end{centering}
\noindent \centering{}\caption{Phonon dispersion curves at 0 K computed with the PINN Al potential
(magenta curves) in comparison with DFT calculations (green curves)
and experimental data \citep{Stedman:1996aa} measured at 80 K (open
blue circles) and 300 K (filled red circles).\label{fig:Phonons}}
\end{figure}

Table \ref{table:al_prop} shows that the potential predicts the equilibrium
lattice constant $a_{0}$ and the elastic constants $c_{ij}$ of FCC
Al in good agreement with DFT values. The potential also demonstrates
reasonable agreement with experimental phonon dispersion curves (Fig.~\ref{fig:Phonons}).
The phonon calculationsutilized the \textsf{phonopy} package \citep{phonopy}
with input from snapshots of a $8\times8\times8$ periodic cell generated
with the PGMC code. For comparison, the plot also shows the results
of DFT calculations performed in this work. We used the Vienna Ab
Initio Simulation Package (VASP) \citep{Kresse1996,Kresse1999} with
the Perdew, Burke, and Ernzerhof (PBE) density functional \citep{PerdewCVJPSF92,PerdewBE96}.
The calculations used the kinetic energy cutoff of 500 eV, 6 irreducible
$k$-points, a Methfessel-Paxton smearing of order 1, and the smearing
width of 0.2 eV. The \textsf{phonopy} package \citep{phonopy} was
utilized with input from a $4\times4\times4$ conventional supercell
containing 256 atoms with the equilibrium lattice constant of 4.04
Å obtained with the PBE functional. Figure~\ref{fig:Phonons} shows
that the DFT dispersion curves compare well with the PINN calculations.

\begin{figure}[htpb]
\noindent \begin{centering}
\includegraphics[scale=0.35]{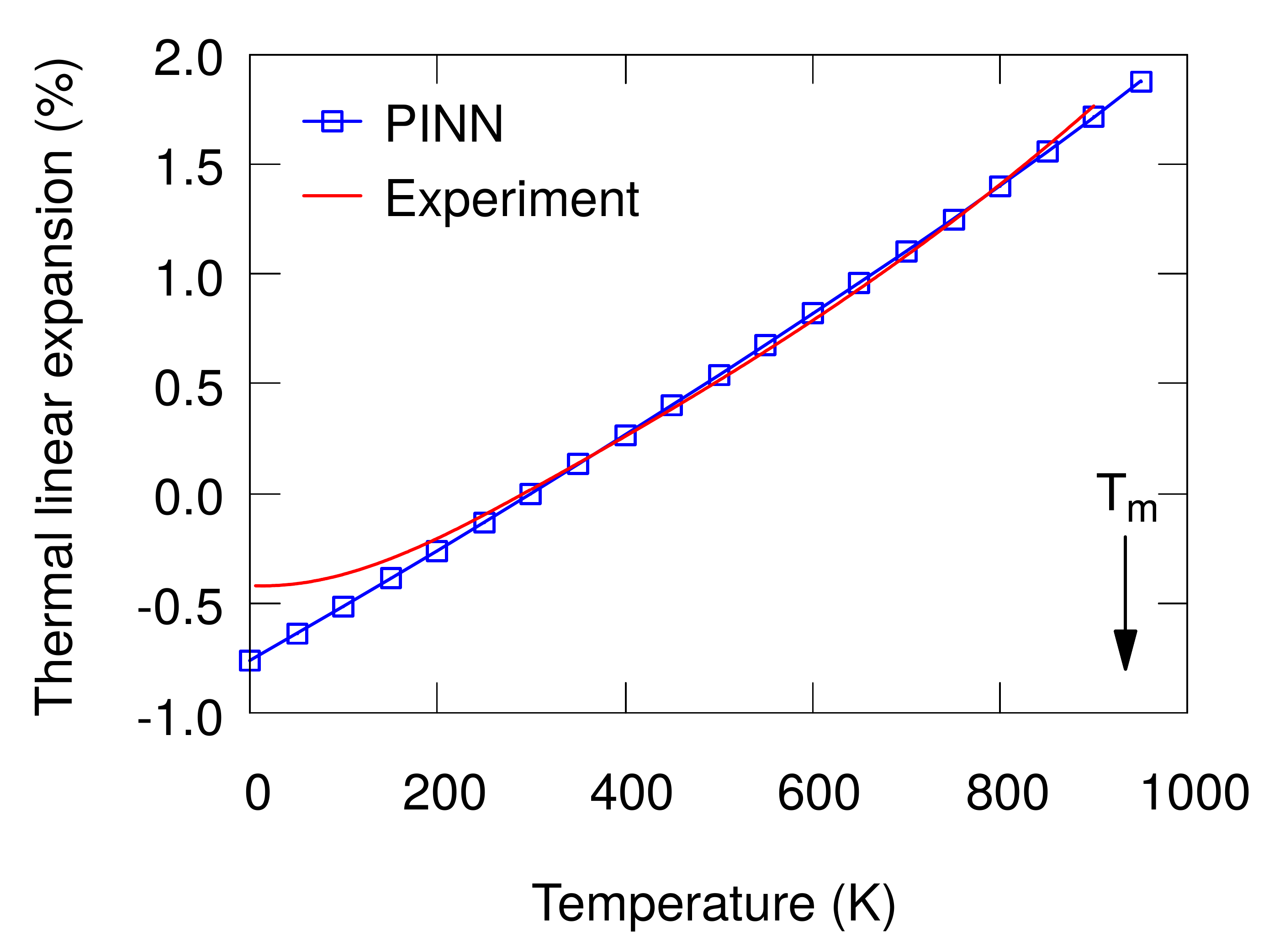}
\par\end{centering}
\noindent \centering{}\caption{Linear thermal expansion coefficient relative to room temperature
predicted by the PINN Al potential in comparison with experiment (the
recommended equation approximating the experimental data \citep{Expansion}).
The arrow indicates the experimental melting temperature.\label{fig:Thermal-expansion}}
\end{figure}

Linear thermal expansion coefficients (relative to 295 K) were computed
by MD simulations on a periodic cubic block containing 10,976 atoms.
The results compare well with experimental data between 295 K and
the melting point (Fig.~\ref{fig:Thermal-expansion}). Deviations
at low temperatures are due to quantum effects that cannot be captured
by a classical potential.

\begin{figure}[htpb]
\noindent \begin{centering}
\includegraphics[scale=0.4]{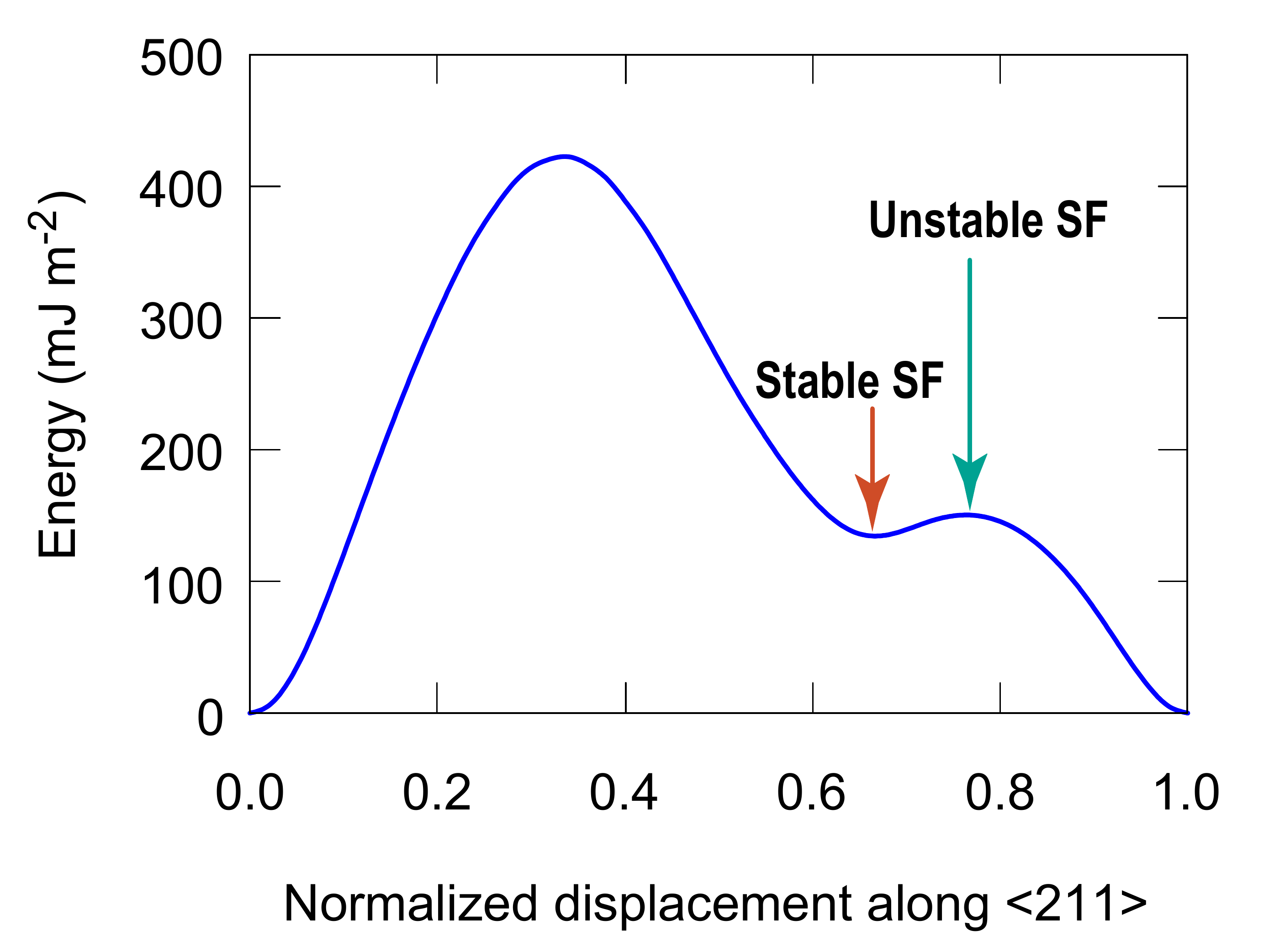}
\par\end{centering}
\noindent \centering{}\caption{Cross-section of the gamma-surface of Al on the (111) plane computed
with the PINN potential. The half-crystal above the (111) plane was
incrementally displaced in the {[}211{]} direction and the energy
was minimized with respect to {[}111{]} atomic displacements after
each increment. The excess energy is plotted against the displacement
normalized by the period of energy in the {[}211{]} direction. The
displacements corresponding to the stable and unstable stacking faults
are indicated. The respective fault energies $\gamma_{SF}$ and $\gamma_{us}$
are indicated in Table \ref{table:al_prop}. \label{fig:SF}}
\end{figure}

Lattice defect energies in Al are also predicted accurately (Table
\ref{table:al_prop}). The surface energies match the DFT values from
the literature. Self-interstitial atoms in Al can be localized in
octahedral or tetrahedral sites, or form split dumbbell configurations.
The potential correctly predicts that the $\left\langle 100\right\rangle $
dumbbell is the most stable configuration. The vacancy migration energy
was computed by the nudged elastic band method \citep{Jonsson98,HenkelmanJ00}
and is well within the bracket of the available DFT values. Given
that the potential accurately reproduces the point defect energies,
it should be suitable for simulations of diffusion, radiation defects,
and similar phenomena mediated by point defect energetics and dynamics.
The stable and unstable stacking fault energies are in good agreement
with DFT data, which is important for simulations of dislocations
and grain boundaries. Fig.~\ref{fig:SF} shows the relevant section
of the gamma-surface on the (111) plane, indicating the stable and
unstable stacking fault positions.

\begin{figure}[htpb]
\noindent \begin{centering}
\textbf{(a)} \includegraphics[scale=0.35]{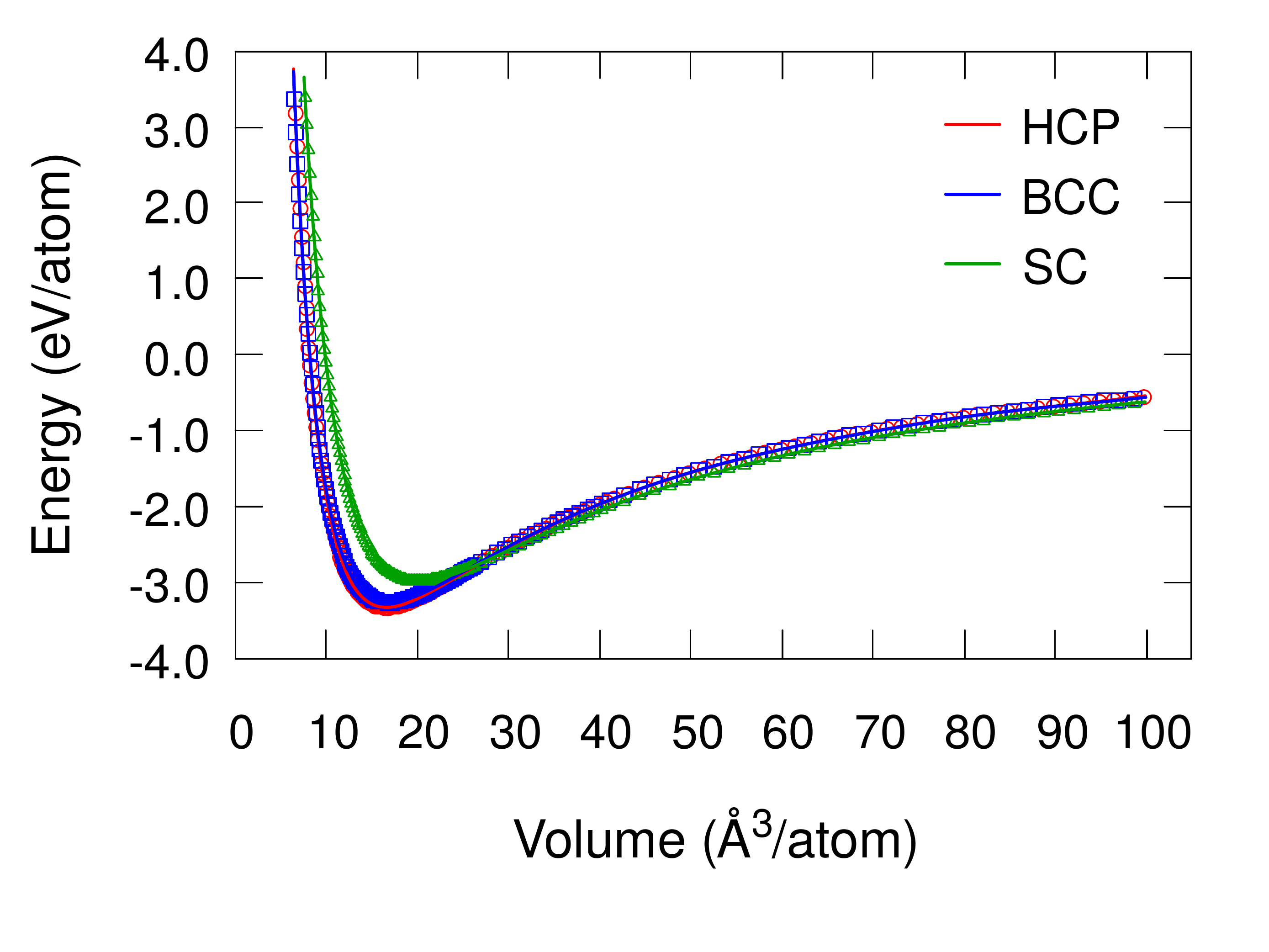}
\par\end{centering}
\bigskip{}

\bigskip{}

\noindent \begin{centering}
\textbf{(b)} \includegraphics[scale=0.35]{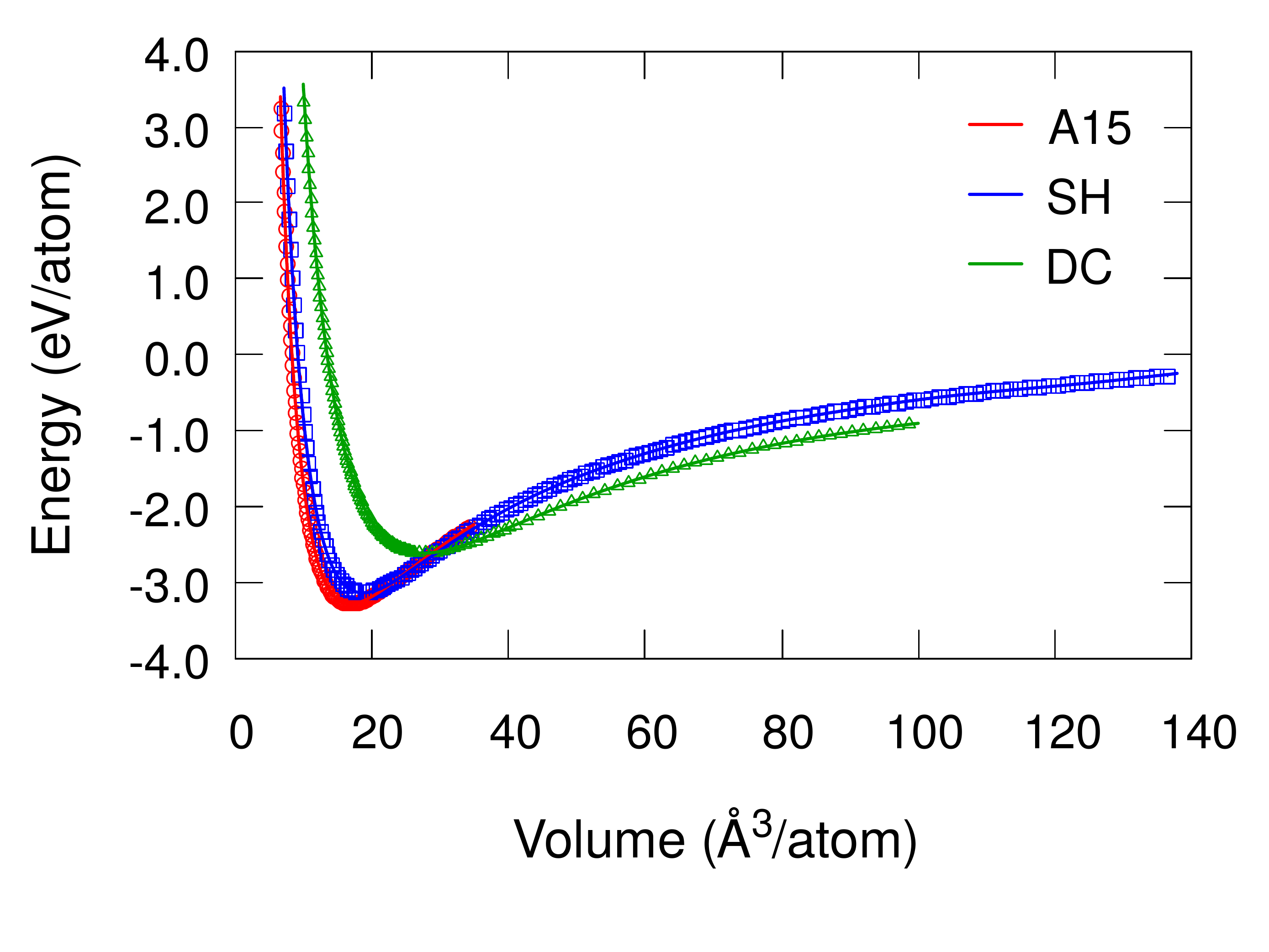}
\par\end{centering}
\noindent \centering{}\bigskip{}
\caption{Energy-volume relations for alternate Al structures computed with
the PINN potential (lines) in comparison with DFT calculations (points).
(a) Hexagonal close-packed (HCP), body-centered cubic (BCC), and simple
cubic (SC) structures. (b) A15 (Cr$_{3}$Si prototype), simple hexagonal
(SH), and diamond cubic (DC) structures.\label{fig:EOS-PINN}}
\end{figure}

\begin{figure}[htpb]
\noindent \begin{centering}
\includegraphics[scale=0.35]{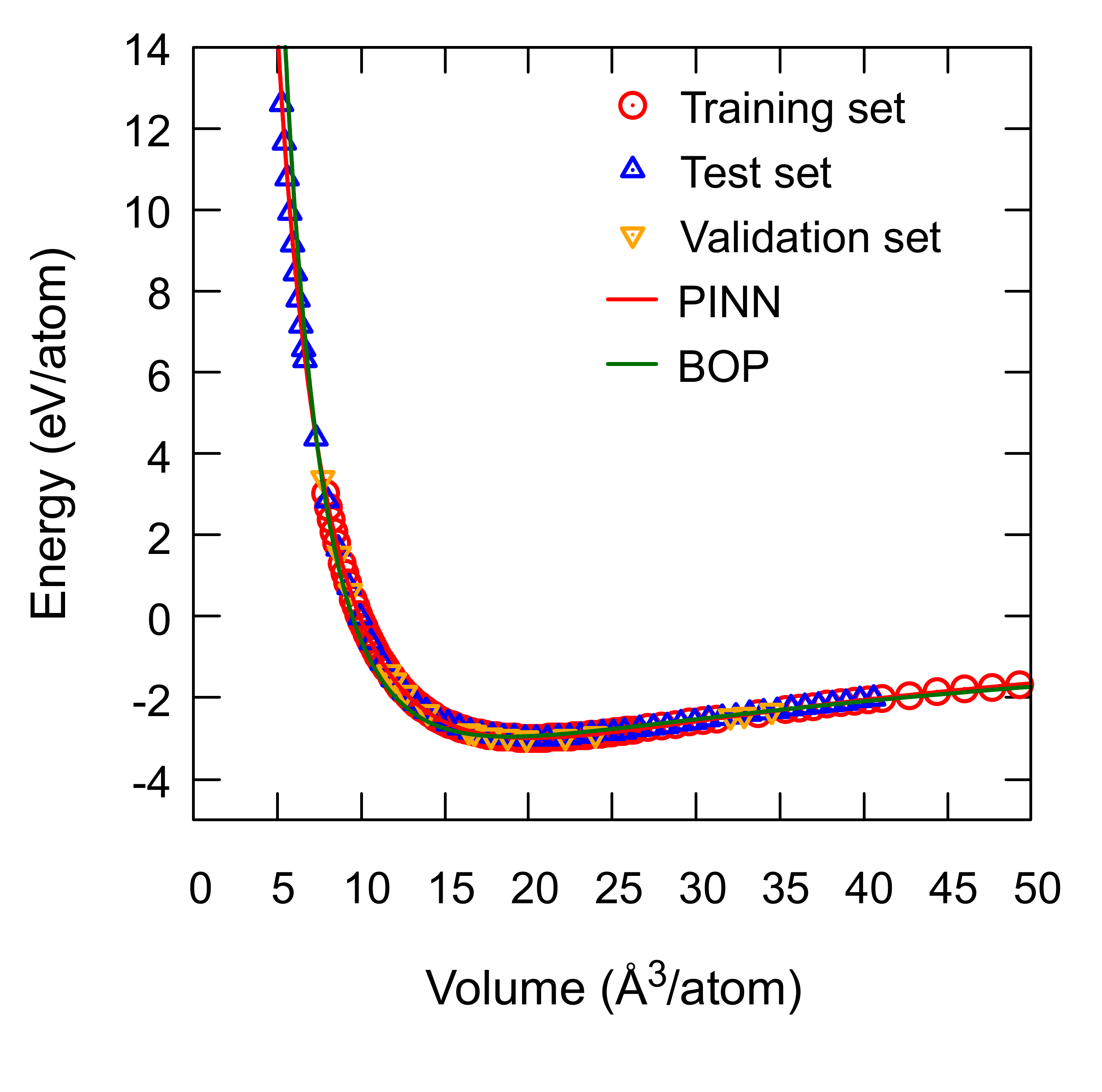}
\par\end{centering}
\noindent \centering{}\caption{Energy-volume relation for simple cubic Al under strong compression
predicted by the PINN and global BOP potentials in comparison with
DFT calculations. The triangular symbols represent DFT energies that
were not used during the training and validation of the potentials.
\label{fig:SC-Compression}}
\end{figure}

Crystal structures other than FCC are also reproduced with high accuracy.
In the interval of atomic volumes sampled by the training database,
the PINN and DFT energy-volume relations are practically indistinguishable
from each other (Fig.~\ref{fig:EOS-PINN}). Importantly, the agreement
continues to be accurate outside the trained volumes. As an example,
Fig.~\ref{fig:SC-Compression} examines the energy-volume curves
for the simple-cubic structure under strong compression. The PINN
potential continues to predict energies that closely match the DFT
points that were \emph{not} included in the training and validation
database. This behavior was observed for all crystal structures tested
in this work. Note that the global BOP potential also extrapolates
well to atomic volumes that were not used during the training and
validation. As discussed in Section \ref{sec:Methodology}, the transferability
of the PINN potential owes its origin to the guidance provided by
the BOP potential.

\begin{figure}[htpb]
\noindent \begin{centering}
\textbf{(a)} \includegraphics[scale=0.35]{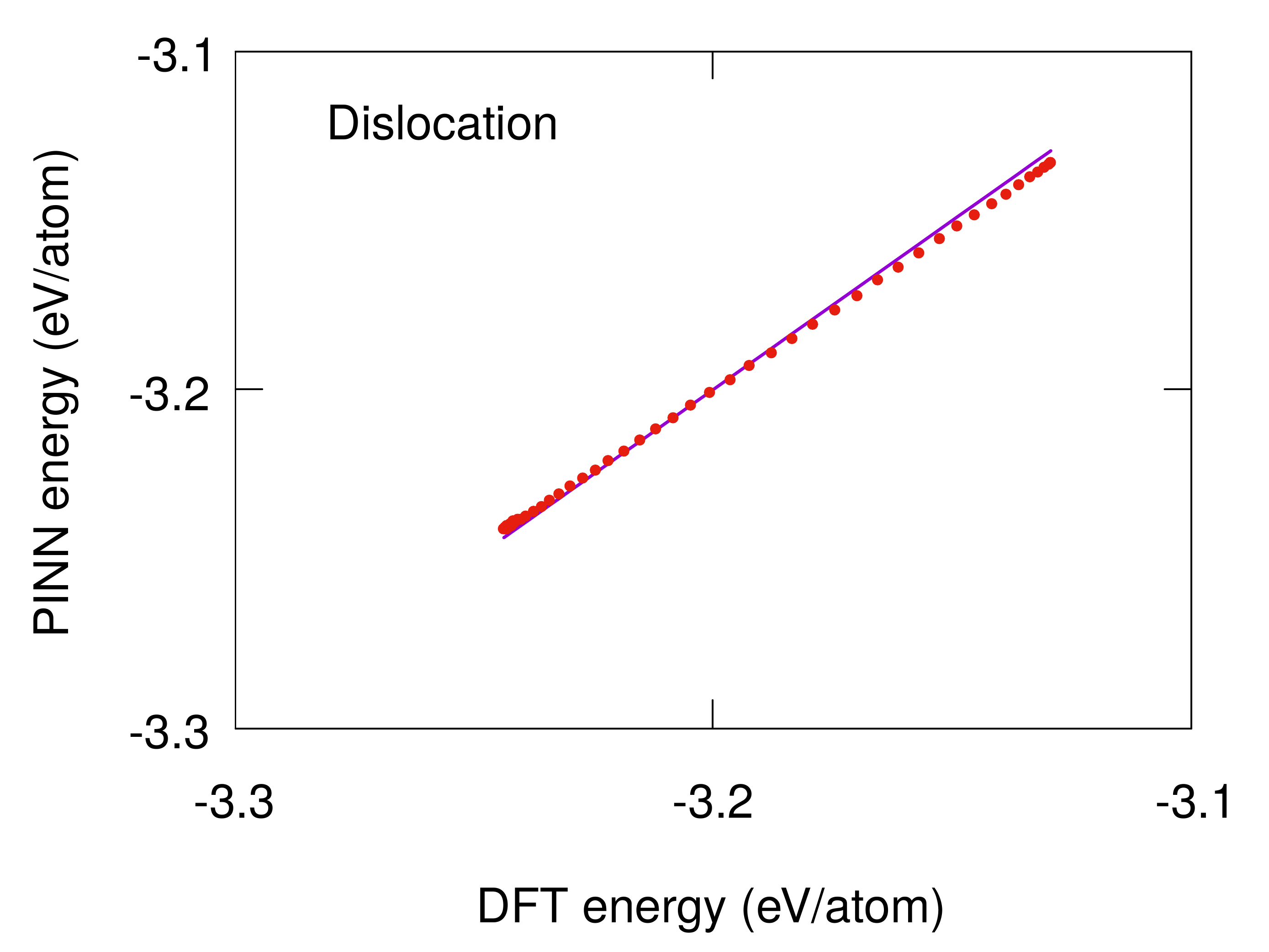}
\par\end{centering}
\bigskip{}
\bigskip{}

\noindent \begin{centering}
\textbf{(b)} \includegraphics[angle=90,scale=0.45]{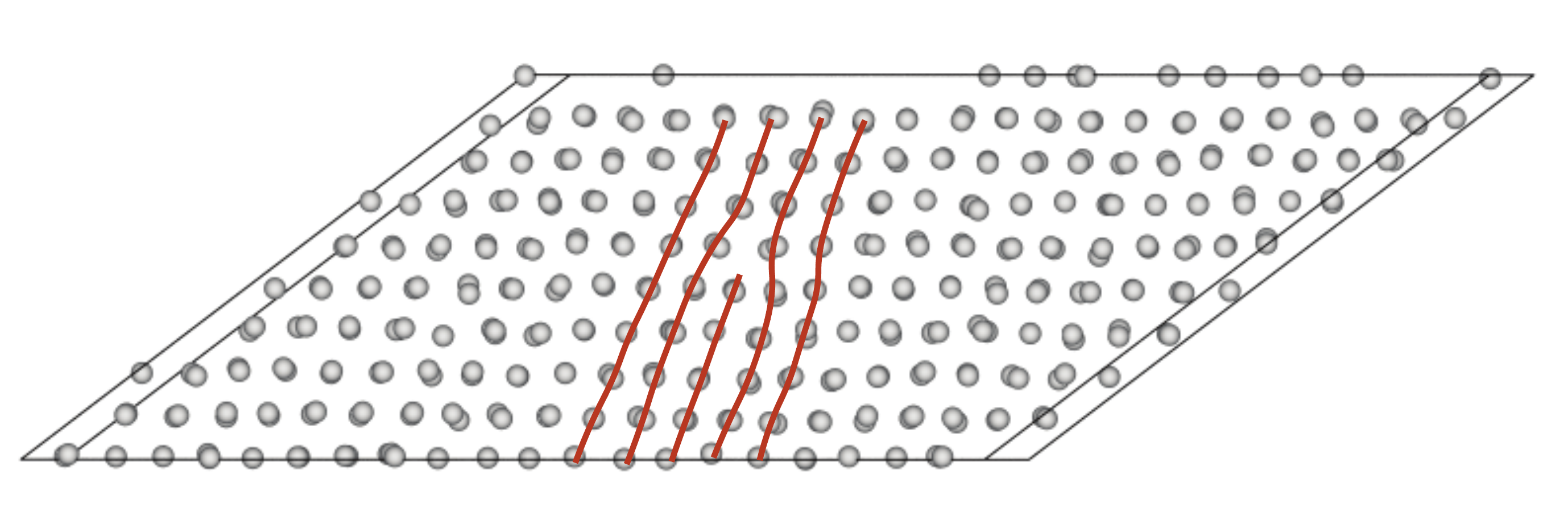}
\par\end{centering}
\noindent \centering{}\caption{(a) Energy computed with the PINN potential compared with the DFT
energy for an edge dislocation in Al at the temperature of 700 K.
The straight line represents perfect fit. (b) Supercell containing
the edge dislocation viewed along the {[}211{]} direction. Selected
crystal plans are traced to show the termination of an extra plane.
\label{fig:Test-dislocation}}
\end{figure}

\begin{figure}[htpb]
\noindent \begin{centering}
\textbf{(a)} \includegraphics[scale=0.35]{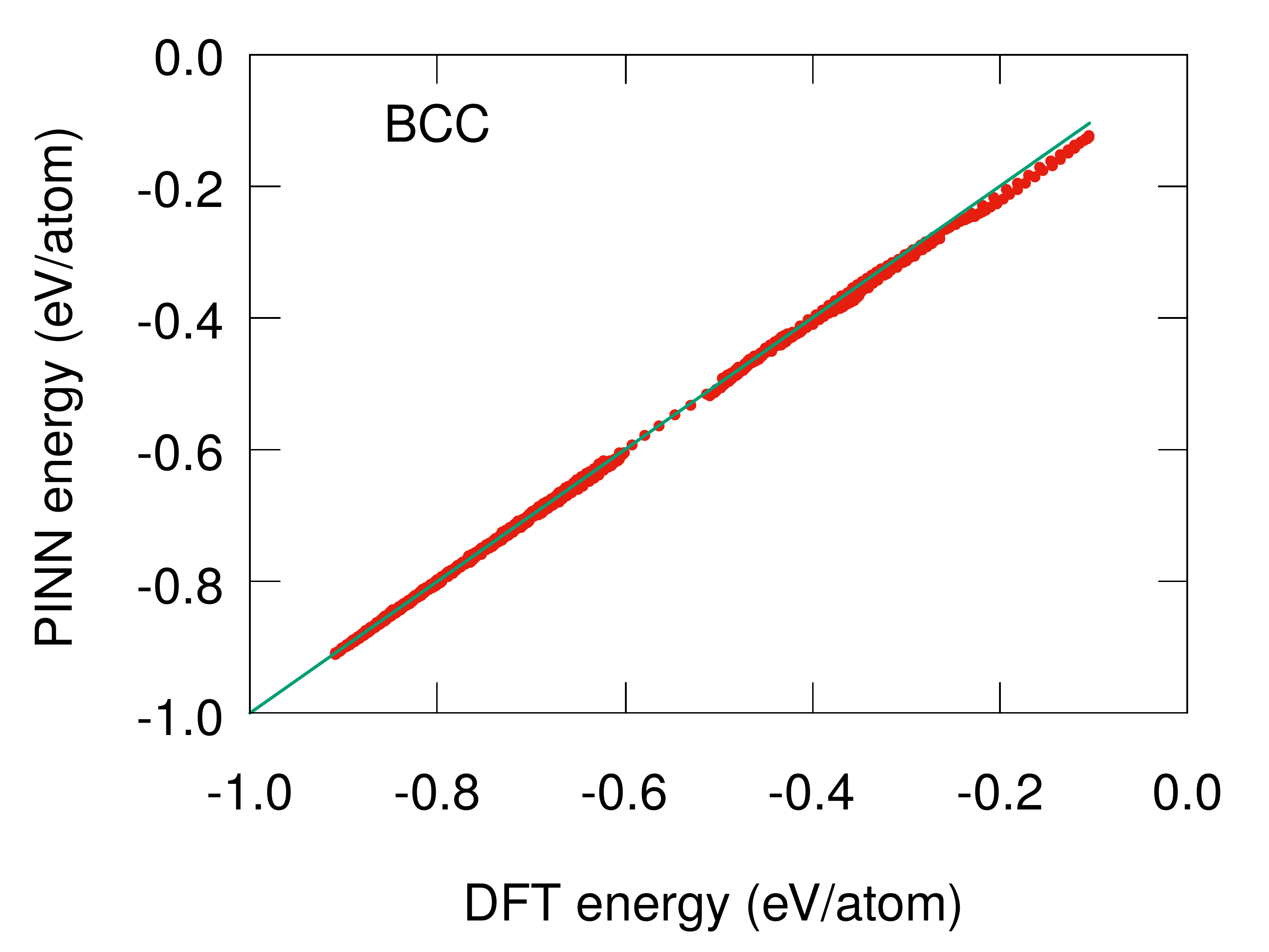}
\par\end{centering}
\bigskip{}

\noindent \begin{centering}
\textbf{(b)} \includegraphics[scale=0.35]{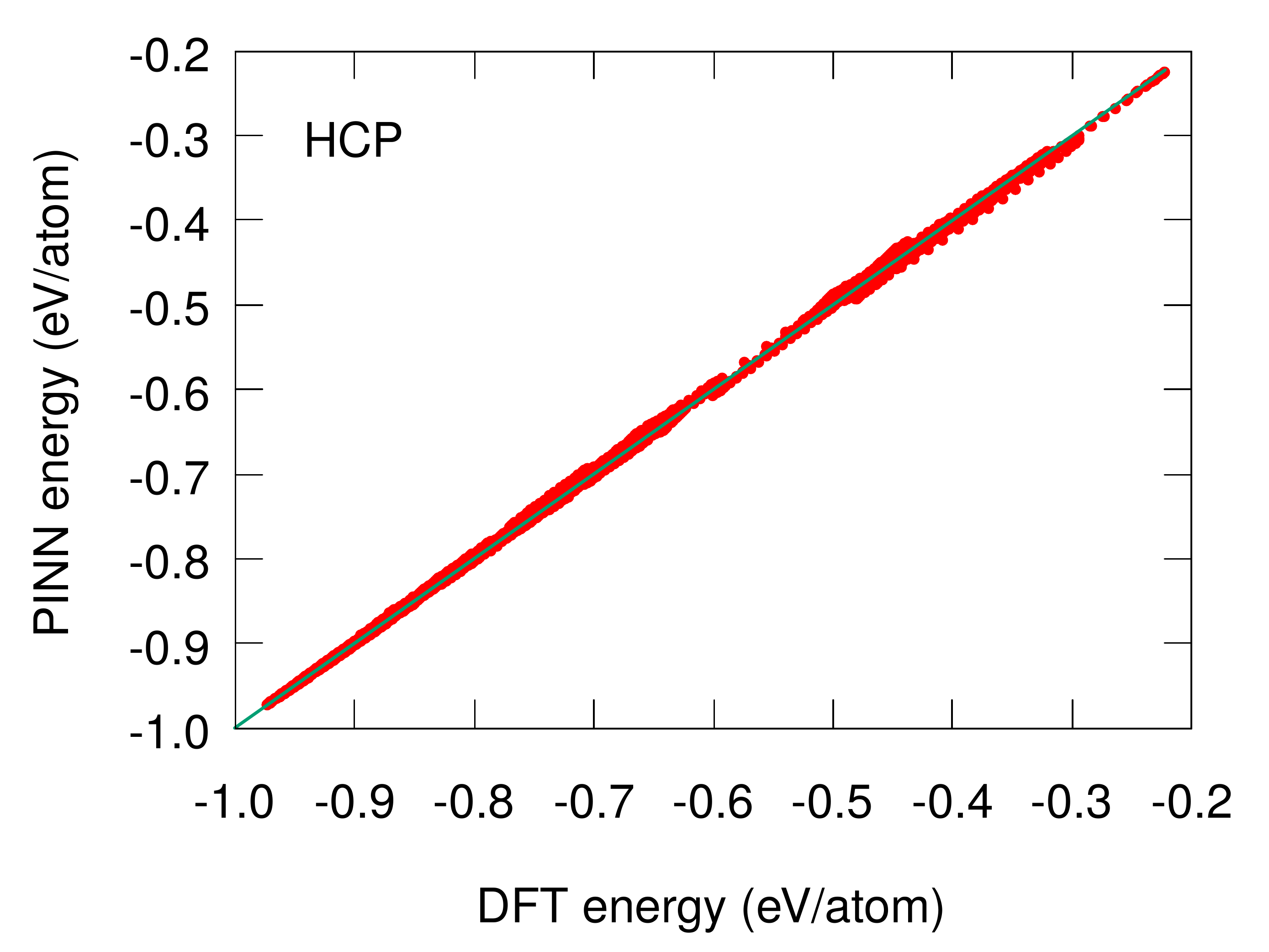}
\par\end{centering}
\noindent \centering{}\caption{Energy computed with the PINN potential compared with DFT energy for
snapshots of NVT (constant temperature and volume) MD simulations
of (a) BCC Al at 1000 K, 2000 K and 4000 K, and (b) HCP Al at 1000
K and 4000 K. The straight line represents perfect fit. \label{fig:Test-examples-1}}
\end{figure}

Testing of a potential is an important step that demonstrates its
scope of applications. The PINN potential was extensively tested for
the ability to reproduce energies of various structures that were
not exposed during the training and validation. As mentioned earlier,
the testing DFT database $\mathcal{T}$ was in fact larger than the
database $\mathcal{F}$$\cup(\cup_{i=1}^{10}\mathcal{V}_{i})$ used
for the training and validation. The agreement between the potential
predictions and the DFT energies was invariably very good. Examples
are shown in Fig.~\ref{fig:Test-dislocation} for a dislocation in
Al and in Fig.~\ref{fig:Test-examples-1} for DFT MD simulations
of BCC and HCP structures at three temperatures exceeding the melting
point. Due to the small supercell size and the periodic boundary conditions,
these crystalline structures were strongly distorted but did not melt
even at 4000 K. Note that the training/validation database only included
these structures at 0 K. Thus the comparison in Fig.~\ref{fig:Test-examples-1}
demonstrates the ability of the potential to extrapolate the energy
outside the training domain. More tests involving both energies and
forces can be found in the Supplementary File.

\section{Further tests and applications\label{sec:Tests}}

\subsection{Calculation of the melting point\label{subsec:Tm}}

In this Section and the subsequent Sections \ref{subsec:Interfaces}
and \ref{subsec:Liquid-structure}, the PINN potential will be used
to investigate the structure and properties of liquid Al and the solid-liquid
coexistence. The motivation for studying systems containing liquid
Al in such detail is twofold. Firstly, the knowledge of liquid properties
is important for the modeling of technological processes such as alloy
casting and additive manufacturing. Secondly, this offers us an opportunity
to assess the transferability of the potential to unknown atomic environments.
Indeed, the bulk liquid phase was \emph{not} represented during the
training and validation. Almost all structures used during the training
and validation were atomically ordered. Some structures were perfectly
ordered, others were strongly distorted, but they still maintained
a significant degree of long-range order. The only exceptions were
the 5\,Å (42 atoms) and 6.5\,Å (79 atoms) isolated clusters at 1200
K and the Wulff-shape (79 atoms) isolated cluster at 2000 K. These
clusters had fully disordered, liquid-like structures. In addition,
a trimer put on the (111) surface at 2000 K caused disordering of
the surface layer in a 103-atom supercell. However, these disordered
structures constituted a small fraction of the database.

\begin{figure}[htpb]
\noindent \begin{centering}
\textbf{(a)} \includegraphics[width=0.65\textwidth]{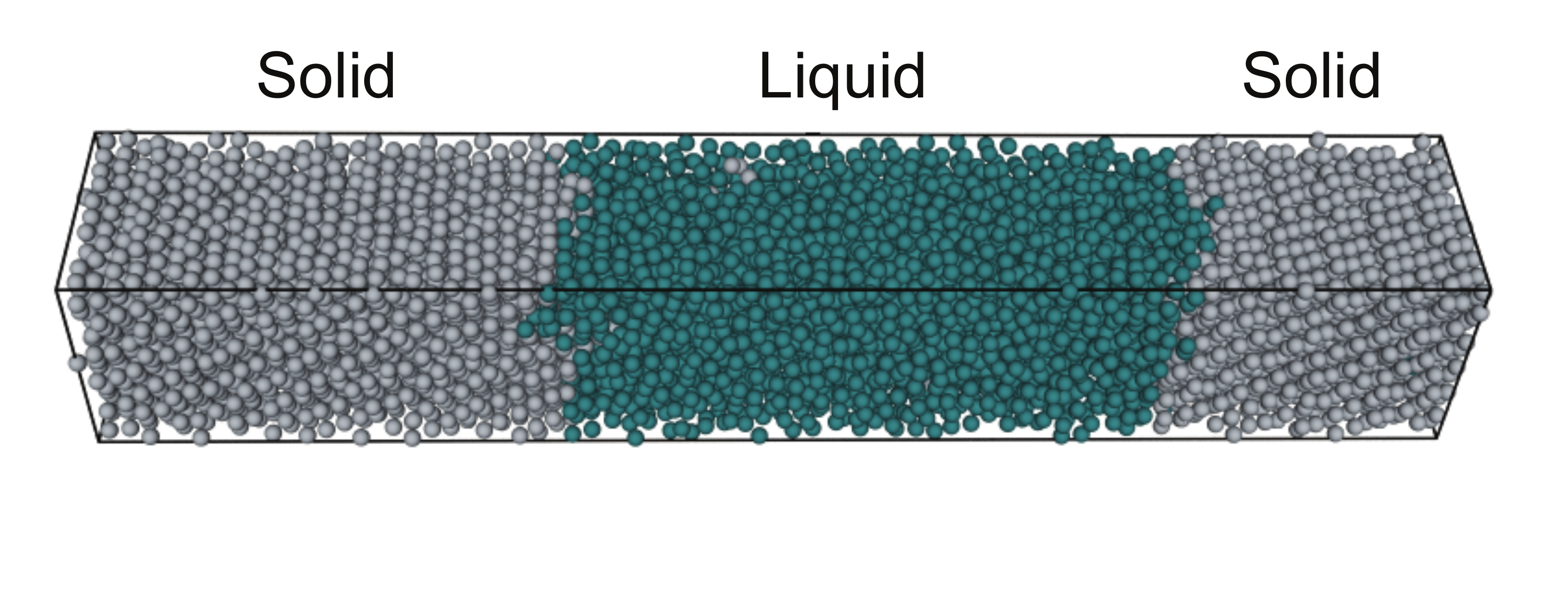}
\par\end{centering}
\bigskip{}

\bigskip{}

\noindent \centering{}\textbf{(b)} \includegraphics[scale=0.32]{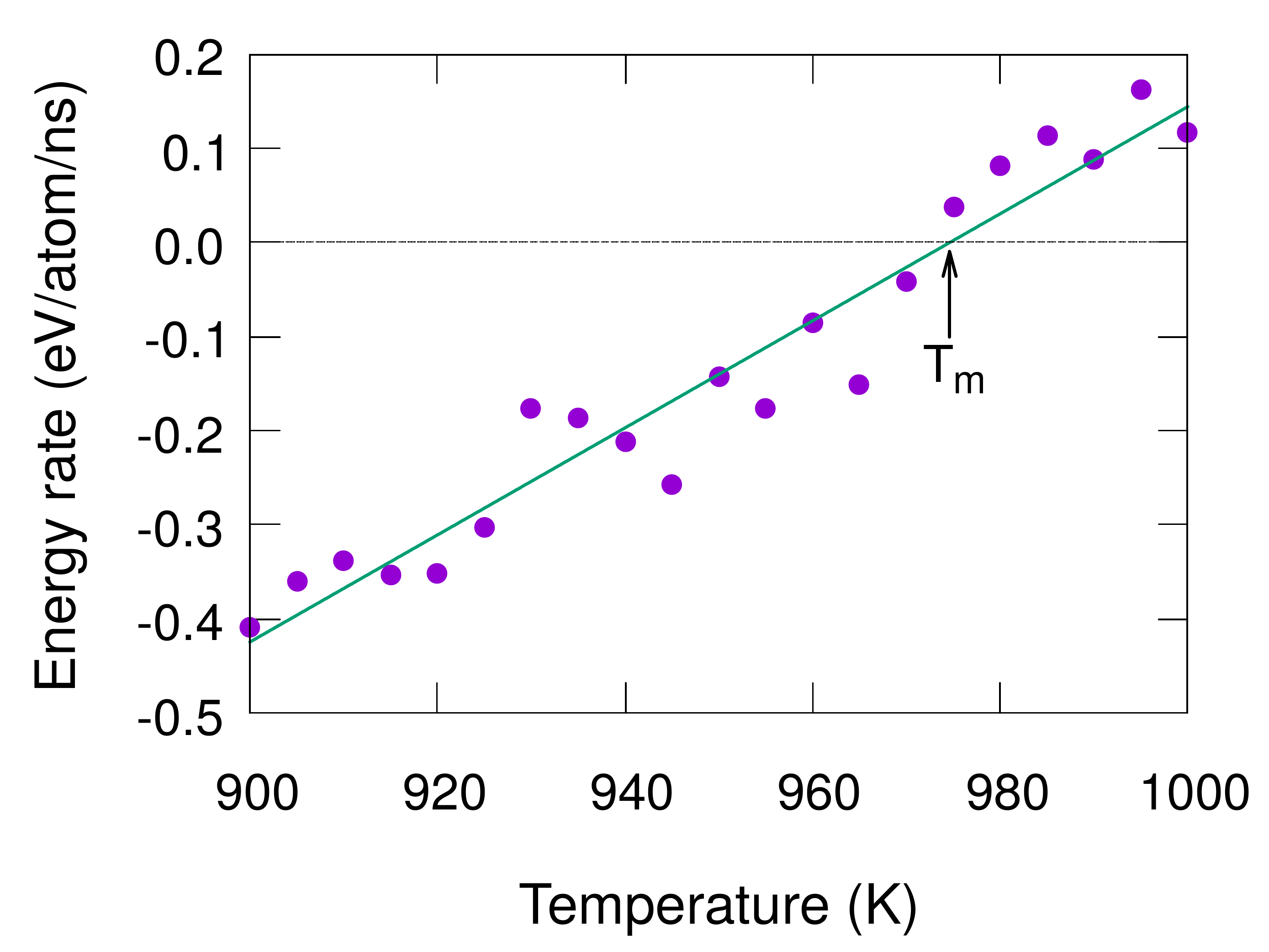}
\caption{(a) Simulation block containing the solid and liquid phases used for
computing the melting temperature. (b) The rate of energy change as
a function of temperature in MD simulations of the solid-liquid system.
The line is a linear fit to the data. The melting point is the temperature
at which the energy rate is zero. \label{fig:melting_temp}}
\end{figure}

The melting temperature $T_{m}$ predicted by the potential was computed
by the interface velocity method described in detail elsewhere \citep{Morris94,Morris02,Pun09b,Purja-Pun:2015aa,Howells:2018aa}.
The simulation block had the dimensions of 29\,Å$\times$30\,Å$\times$185\,Å\ and
contained 9,000 atoms, which were partitioned between the solid and
liquid phases separated by a (111) interface normal to the long direction.
NPT MD simulations (constant temperature and zero pressure) were executed
at a series of temperatures near the expected melting point. During
the simulations, the solid phase was either melting or crystallizing,
depending on whether the chosen temperature happened to be above or
below $T_{m}$. Accordingly, the system energy was either increasing
with time or decreasing. The rate of the energy change could be converted
to the solid-liquid interface velocity to find the temperature at
which the velocity vanished. Instead, it was sufficient to monitor
the energy rate itself and identify the melting point with the temperature
at which this rate was zero. In Fig.~\ref{fig:melting_temp}, the
energy rate is plotted against temperature for several simulation
runs. Interpolation using a linear regression gives $T_{m}=(975\pm3)$
K (the uncertainty indicates one standard deviation). We consider
the agreement with the experimental melting point of Al (933 K \citep{Kittel})
encouraging given that it was achieved without any direct fit.

\subsection{Interface tensions\label{subsec:Interfaces}}

\begin{figure}[htpb]
\noindent \centering{}\includegraphics[width=0.7\textwidth]{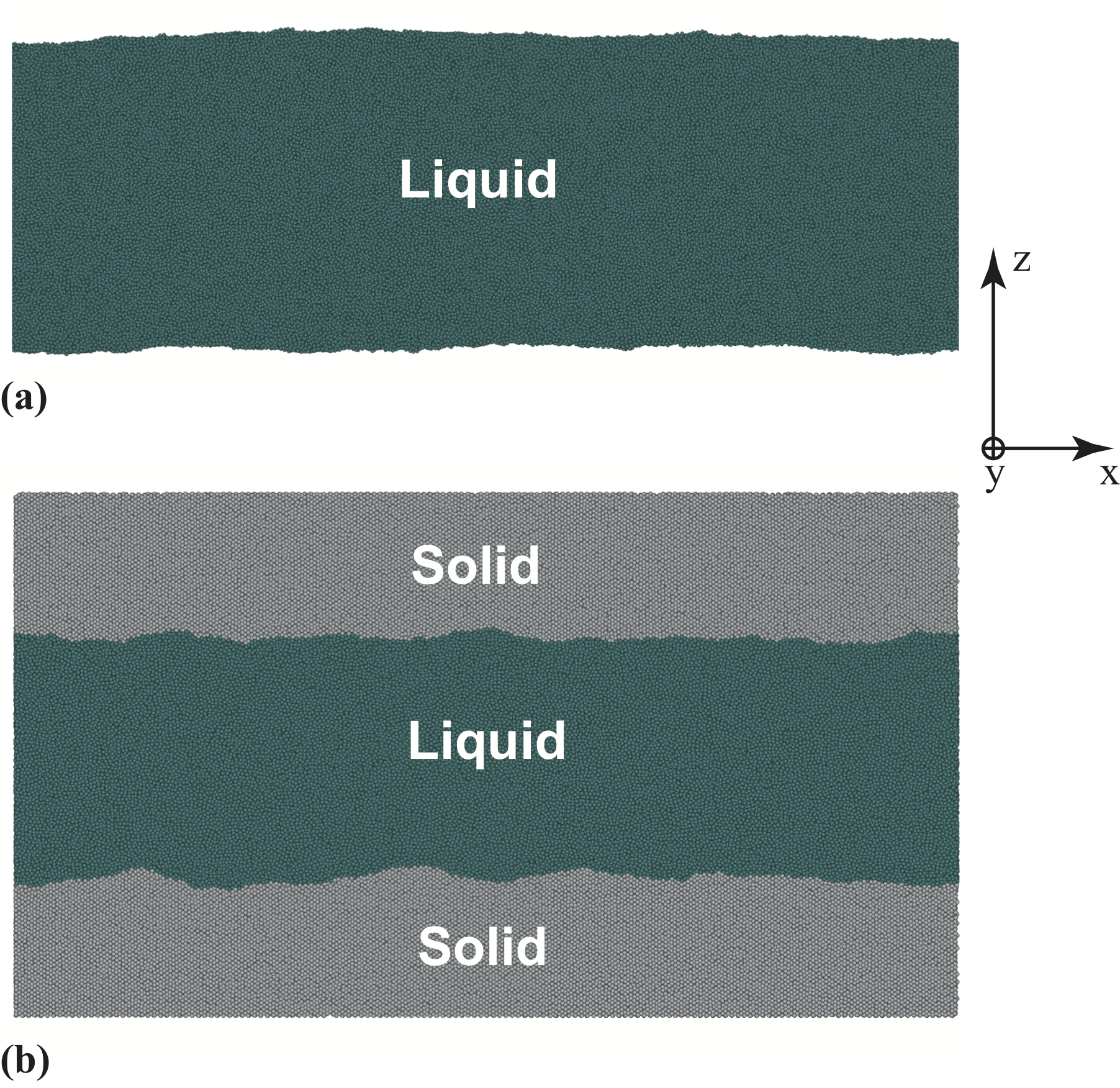}\caption{Simulation blocks used for computing the interface tensions in Al.
(a) Liquid film with open surfaces. Periodic boundary conditions are
applied in the $x$ and $y$ directions. (b) Solid-liquid coexistence
system. The crystallographic directions $\langle100\rangle$ and $\langle110\rangle$
in the solid phase are parallel to the $x$ and $y$ axes, respectively.
Periodic boundary conditions are applied in all three directions.
The images are visualized using the potential energies of atoms. \label{fig:interfaces}}
\end{figure}

The liquid and solid-liquid interface tensions in Al were computed
by the capillary fluctuation method \citep{Hoyt01,Morris02a,Rozas2011,Karma-2012a,Mishin2014,Asadi2015}.
In this method, the interface is aligned normal to the $z$-direction
and has a ribbon-like shape with the $y$-dimension $w$ much smaller
than the $x$-dimension $l$. Periodic boundary conditions are imposed
in the $(x,y)$ plane. An example is shown in Fig.~\ref{fig:interfaces}a
for liquid surfaces of a free-standing film with $l=622$ Å and $w=29$
Å.

Capillary fluctuations manifest themselves in stochastic variations
of the interface shape $z(x)$, which can be quantified by the function
$h(x)=z(x)-\overline{z}$, where $\overline{z}$ is the average interface
position. Fourier transformation of $h(x)$ gives the power spectrum
$\vert A(k)\vert^{2}$ of the capillary waves, $A(k)$ being the Fourier
amplitude and $k$ the wave number. The canonical expectation value
$\left\langle \vert A(k)\vert^{2}\right\rangle $ is obtained by averaging
the power spectrum over multiple snapshots and the two interfaces
present in the system. By fitting $\left\langle \vert A(k)\vert^{2}\right\rangle $
with the function \citep{Hoyt01,Morris02a,Rozas2011,Karma-2012a,Mishin2014,Asadi2015}

\begin{equation}
\langle\vert A(k)\vert^{2}\rangle=\dfrac{k_{B}\,T}{lw\,(\gamma+\gamma^{\prime\prime})\,k^{2}},\label{eq:CW}
\end{equation}
the interface stiffness $(\gamma+\gamma^{\prime\prime})$ can be extracted.
Here $\gamma$ is the interface tension, $\gamma^{\prime\prime}$
is the second derivative of $\gamma$ with respect to the inclination
angle, and $k_{B}$ is Boltzmann's constant. In practice, $(\gamma+\gamma^{\prime\prime})$
is obtained from the slope of the plot $\left\langle \vert A(k)\vert^{2}\right\rangle ^{-1}$
versus $k^{2}$ in the long-wave (small $k$) limit.

\begin{figure}[htpb]
\noindent \begin{centering}
\textbf{(a)} \includegraphics[width=0.55\textwidth]{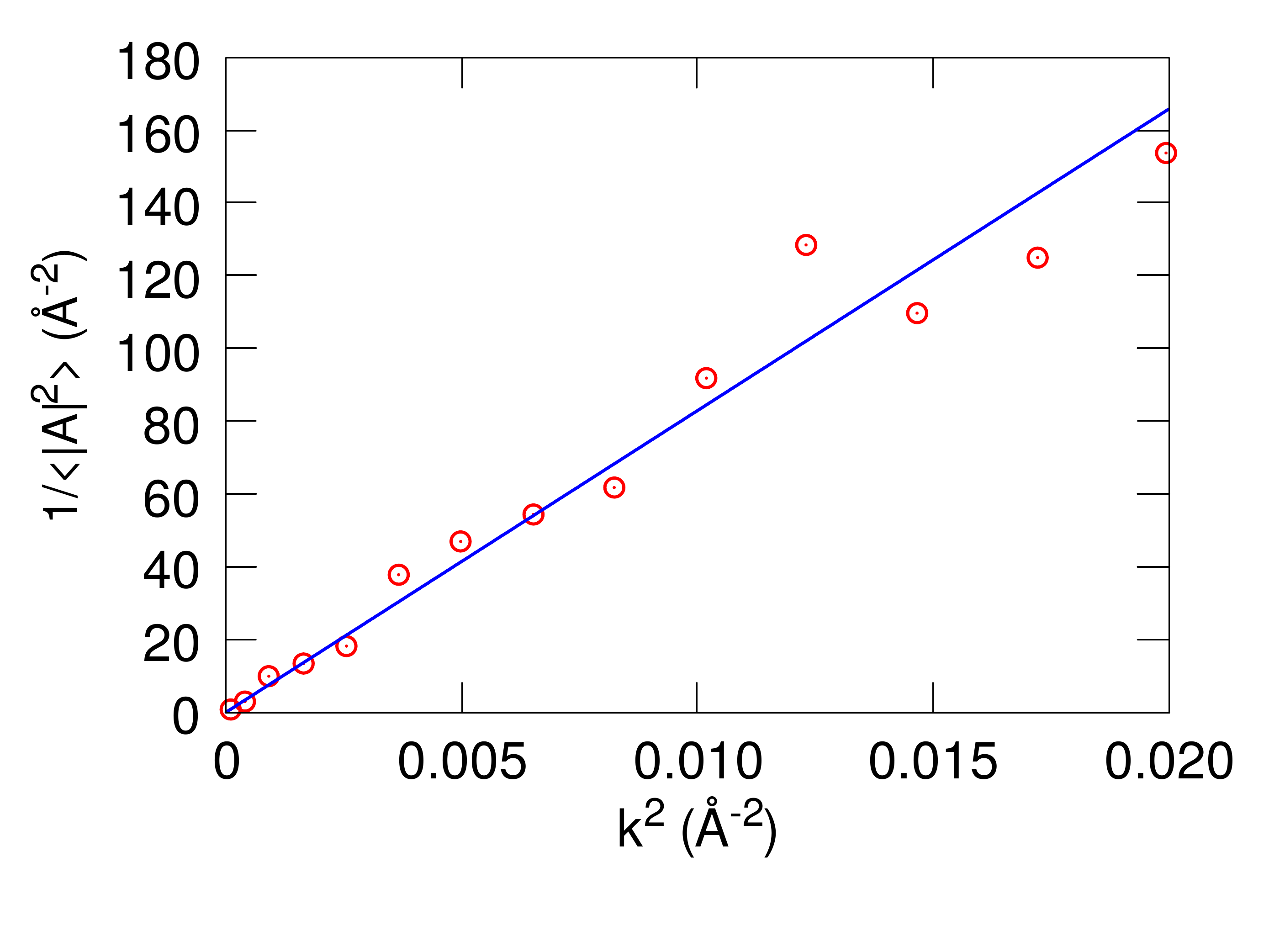}
\par\end{centering}
\bigskip{}

\bigskip{}

\noindent \centering{}\textbf{(b)} \includegraphics[width=0.55\textwidth]{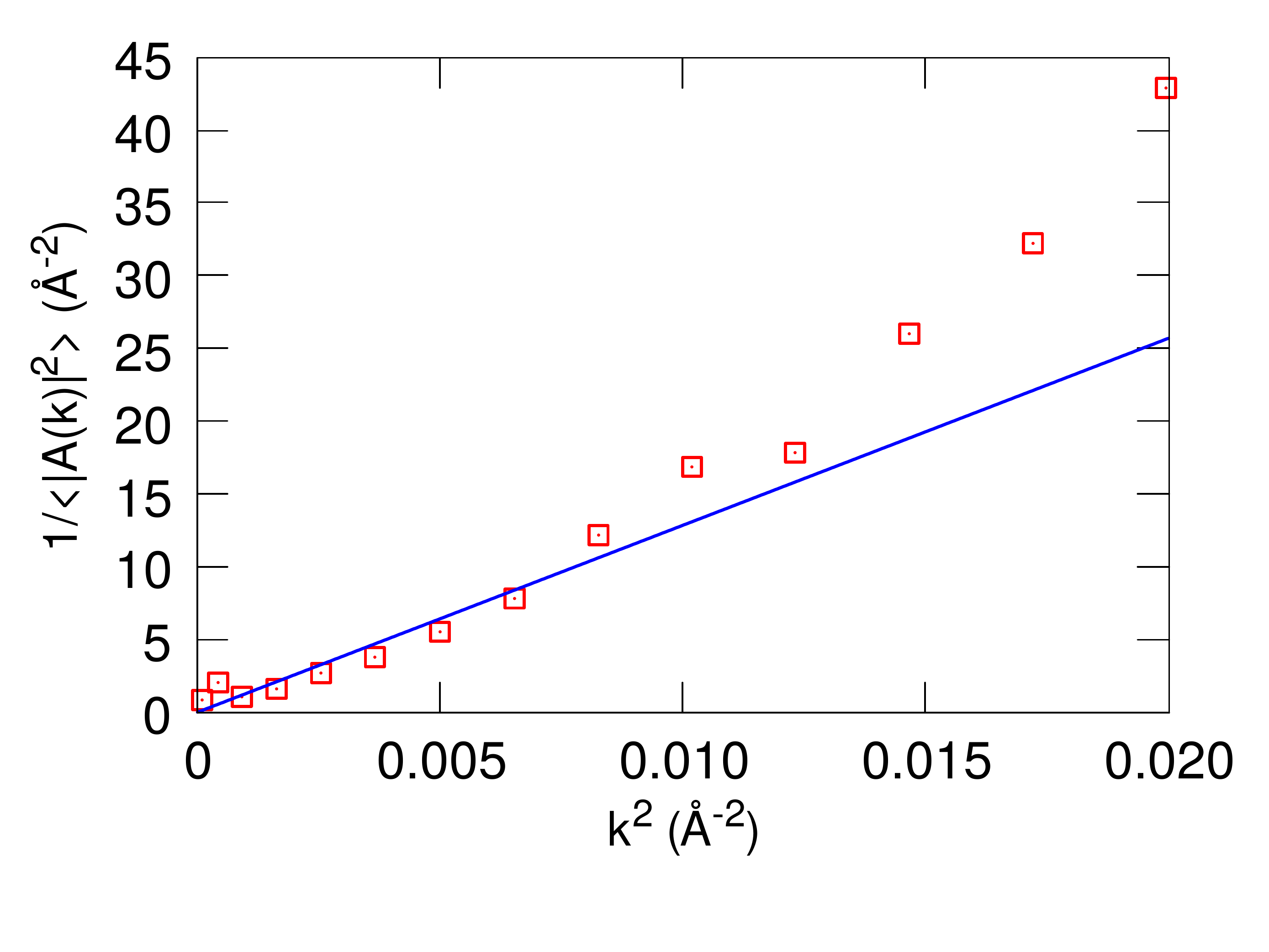}
\caption{Inverse power of capillary waves versus the wave number squared for
(a) liquid Al surface and (b) Al solid-liquid interface computed with
the PINN potential. The lines represent linear fits in the long-wave
limit. \label{fig:CW-2}}
\end{figure}

For computing the liquid surface tension, the simulation block (Fig.~\ref{fig:interfaces}a)
was equilibrated at the melting temperature (975 K) followed by a
0.57 ns long NVE MD production run (constant volume and energy). Snapshots
containing the atomic coordinates were saved every 10 ps. At the post-processing
stage, each snapshot was divided into 200 thin bins normal to the
$x$ axis. The upper and lower interface positions were found by averaging
10 largest (respectively, 10 smallest) $z$-coordinates of atoms in
the bin. Because the interfaces are not atomically sharp, the averaging
is more appropriate than simply taking the largest and smallest coordinates.
The power spectrum of the capillary waves was obtained by a discrete
Fourier transformation of the interface locations in the bins. For
a liquid surface $\gamma^{\prime\prime}=0$. Linear fit to the $\left\langle \vert A(k)\vert^{2}\right\rangle ^{-1}$
versus $k^{2}$ plot in the $k\rightarrow0$ limit (Fig.~\ref{fig:CW-2}a)
gives the surface tension of $\gamma=0.610$ J\,m$^{-2}$.

To verify this result, another, independent method was applied. Namely,
the thin film in Fig.~\ref{fig:interfaces}a is subject to the Laplace
pressure $p=2\gamma/d$, where $d=196$ Å is the film width in the
$z$-direction. The pressure in the inner region of the film unaffected
by the surfaces was computed by averaging over the entire set of snapshots.
Knowing the pressure, we obtain $\gamma=pd/2=0.613$ J\,m$^{-2}$
(Table \ref{table:surface_tension}). This number is close to the
previous result, which lends credence to the capillary wave methodology
used in this work. It should be emphasized that the Laplace pressure
is mechanical in nature and is caused by the interface stress, whereas
the stiffness appearing in the capillary fluctuation method is related
to the interface free energy $\gamma$ (which we refer to here as
tension). While the interface stress and interface free energy are
conceptually different properties, they are numerically equal for
a liquid surface. It is this equality that enabled us to compute the
same surface tension by the two different methods. This cannot be
done for the solid-liquid interface discussed below, for which the
interface stress and interface free energy (tension) are generally
different \citep{Willard_Gibbs,Cahn79a,Cammarata1994b,Cammarata08,Frolov09,Frolov2010b}.

For comparison, the same calculations were performed with one of the
widely used EAM Al potentials \citep{Mishin99b}. The melting temperature
predicted by this potential is 1042 K \citep{Ivanov08a}. A larger
simulation block could be afforded thanks to the computational efficiency
of the EAM. The $\left\langle \vert A(k)\vert^{2}\right\rangle ^{-1}$
versus $k^{2}$ plot can be found in the Supplementary File. Although
the EAM potential is expected to be less accurate than the PINN potential,
the surface tensions obtained are reasonably close to the respective
PINN values (Table \ref{table:surface_tension}). For completeness,
Table \ref{table:surface_tension} also cites experimental data \citep{Poirier:1987aa,Egry:2008aa}.
The experimental surface tension tends to be higher than the computed
ones. However, comparison with experiment should be taken with caution.
Experimental measurements are conducted on much larger droplets, typically
several millimeters in radius \citep{Schmitz:2009aa,Zhao:2017aa,Poirier:1987aa,Egry:2008aa}.
The tension is extracted from droplet shape oscillations during electromagnetic
levitation. The accuracy of the results is limited by many factors,
such as temperature control, surface contamination and evaporation.

Solid-liquid interfaces were created in a simulation block containing
both phases in thermodynamic equilibrium with each other at the melting
temperature of the respective potential (Fig.~\ref{fig:interfaces}b).
The interfaces were parallel to the (110) plane of the solid phase,
with periodic boundary conditions imposed in all three directions.
To ensure zero pressure, the lattice constant of the solid phase was
adjusted according to the thermal expansion coefficient at the chosen
temperature. The system size in the $z$-direction was also adjusted
to remove the normal stress. Once thermodynamic equilibrium was reached,
a production run was implemented in the NVE MD ensemble for 0.57\,ns
(PINN) or 10\,ns (EAM). To compute the interface shape, the liquid
phase in each snapshot was ``removed'' by discarding all atoms whose
energy was greater than -3.142 eV. The surfaces of the remaining solid
mimicked the solid-liquid interfaces, which were then binned to determine
the $z$-coordinates of the interfaces positions as discussed above.
The remaining steps were the same as for the liquid surfaces.

Figure \ref{fig:CW-2}b shows the $1/\langle\vert A(k)\vert^{2}\rangle$
versus $k^{2}$ plot computed with the PINN potential (see Supplementary
File for the plot obtained with the EAM potential). The interface
stiffness obtained is 95 mJ\,m$^{-2}$ (Table \ref{table:surface_tension-1}),
which is almost an order of magnitude smaller than the liquid surface
tension. Since the calculations were performed for a single interface
orientation, the interfaces tension $\gamma$ cannot be separated
from the torque term $\gamma^{\prime\prime}$. Thus only the stiffness
values $(\gamma+\gamma^{\prime\prime})$ are reported. The stiffness
predicted by the EAM potential \citep{Mishin99b} is slightly higher
but close to the PINN result (Table \ref{table:surface_tension-1}).
Taken together with the liquid surface results, we observe that the
EAM potential \citep{Mishin99b} tends to overestimate the interface
tensions but otherwise demonstrates reasonable accuracy. Other authors
reported even larger stiffness values using different EAM \citep{Morris02a}
and MEAM \citep{Asadi2015} potentials. Comparison with experiment
is problematic. The solid-liquid interface stiffness has only been
estimated by indirect methods, such as back-calculation from experimentally
observed crystal nucleation rates, measurements of dihedral angles
(this requires the knowledge of other tensions), or melting point
depression \citep{Jiang08}. The anisotropy of stiffness is not taken
into account. Nevertheless, some experimental data is included in
Table \ref{table:surface_tension-1} for completeness.

\subsection{Liquid structure and properties\label{subsec:Liquid-structure}}

We next discuss the structure and properties of liquid Al as a single
phase. Two structural properties will be examined: the radial distribution
function (RDF) and the bond angle distribution.

\begin{figure}[htpb]
\noindent \begin{centering}
\textbf{(a)} \includegraphics[width=0.6\textwidth]{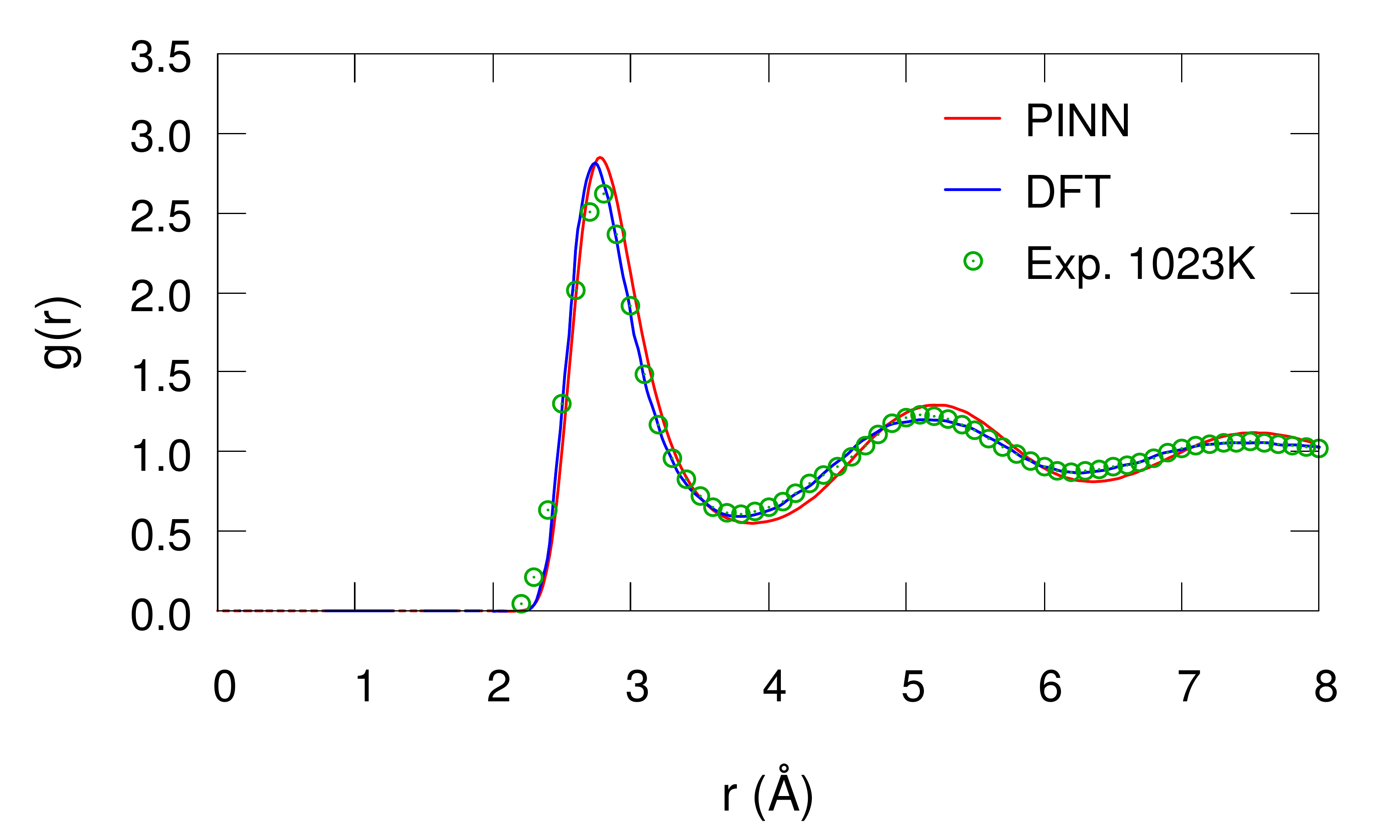}
\par\end{centering}
\bigskip{}

\bigskip{}

\noindent \centering{}\textbf{(b)} \includegraphics[width=0.6\textwidth]{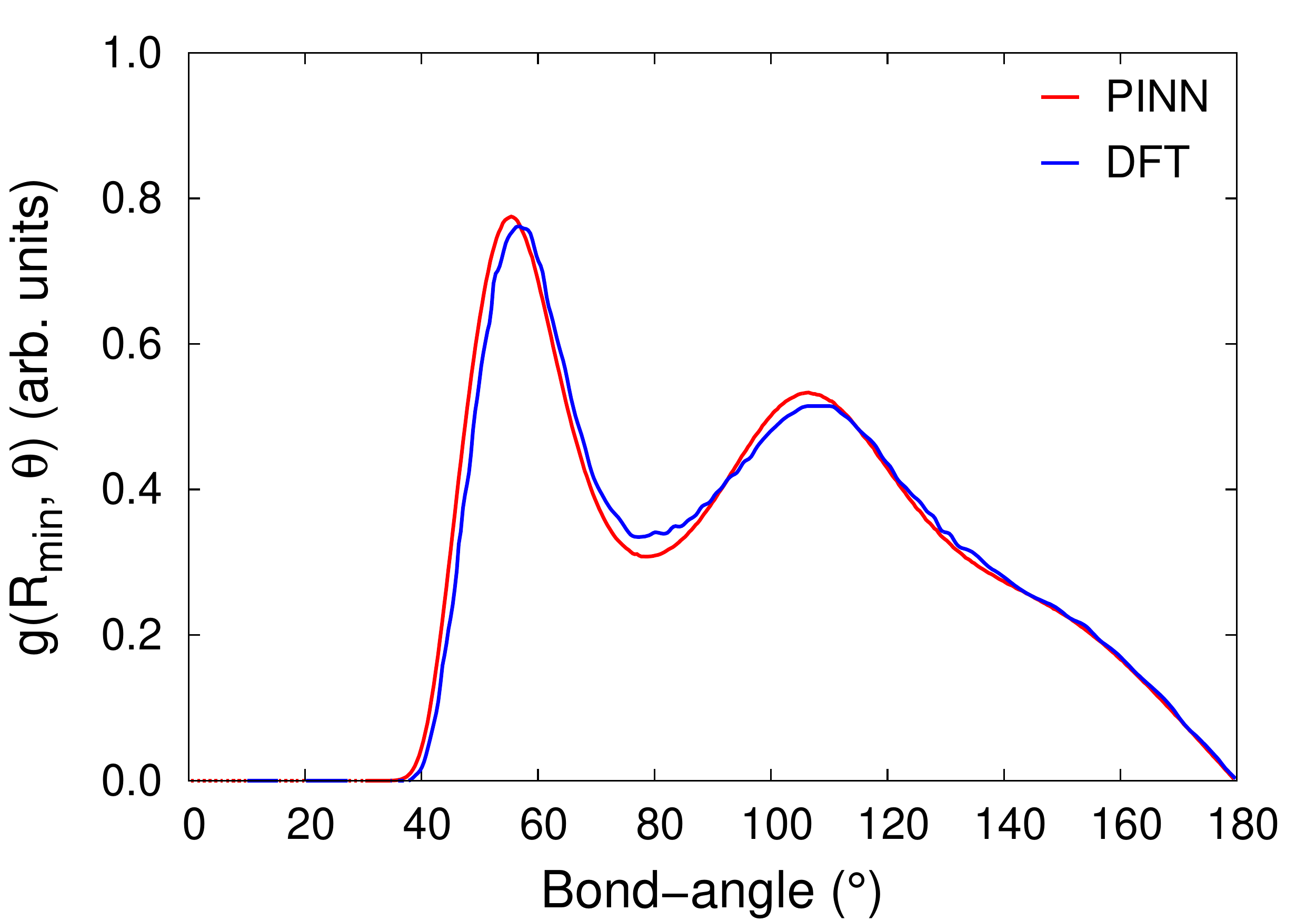}
\caption{Structure of liquid Al at 1000 K predicted by the PINN potential in
comparison with experimental data \citep{Mauro:2011wc,Jakse:2013aa}
and DFT calculations \citep{Alemany:2004,Jakse:2013aa}. (a) Radial
distribution function; (b) bond-angle distribution function. \label{fig:Liquid-structure}}
\end{figure}

The RDF $g(r)$ was computed by averaging over 250 snapshots saved
during a 30 ps NPT MD simulation at several temperatures. The simulation
block contained 10,976 atoms with periodic boundaries. Figure \ref{fig:Liquid-structure}a
shows the RDFs at the temperatures of 1000 K (PINN and DFT) and 1013
K (experiment). Similar plots for the temperatures of 875 K, 1125
K and 1250 K are included in the Supplementary File. In all cases
tested, the results predicted by the PINN potential were in very good
agreement with both experimental data and DFT calculations \citep{Mauro:2011wc,Jakse:2013aa}.

The same MD snapshots were used to compute the distribution function
$g(R_{\textrm{min}},\theta$) of bond angles $\theta$. The bonds
were defined as vectors connecting a chosen atom with its neighbors
lying within the radius $R_{\textrm{min}}$ of the first minimum of
$g(r)$. The function $g(R_{\textrm{min}},\theta$) obtained (Fig.~\ref{fig:Liquid-structure}b)
compares well with the results of DFT MD simulations \citep{Alemany:2004}.

Calculations of density, viscosity and diffusivity were performed
at 10 temperatures ranging from 1050 K to 1500 K at 50 K intervals
using a periodic cubic block containing 32,000 atoms. At each temperature,
the system was equilibrated by an NPT MD simulation for at least 200
ps to ensure decorrelation from the previous temperature and reach
the equilibrium density. The equilibration was followed by 20 to 30
production runs, 100 ps each, implemented in the NPT ensemble for
density calculations and NVT ensemble for computing the viscosity
and diffusivity. The MD integration step was 1 fs. The atomic stress,
atomic positions, velocities, energies, and other relevant parameters
were measured at every MD step. The viscosity and diffusion coefficients
were computed by the Green-Kubo method following the methodology described
in Ref.~\citep{Mondello:1997aa}. As a cross-check, the diffusion
coefficients were also computed from mean-squared atomic displacements
using the Einstein equation.

\begin{figure}[htpb]
\noindent \centering{}\includegraphics[width=0.7\textwidth]{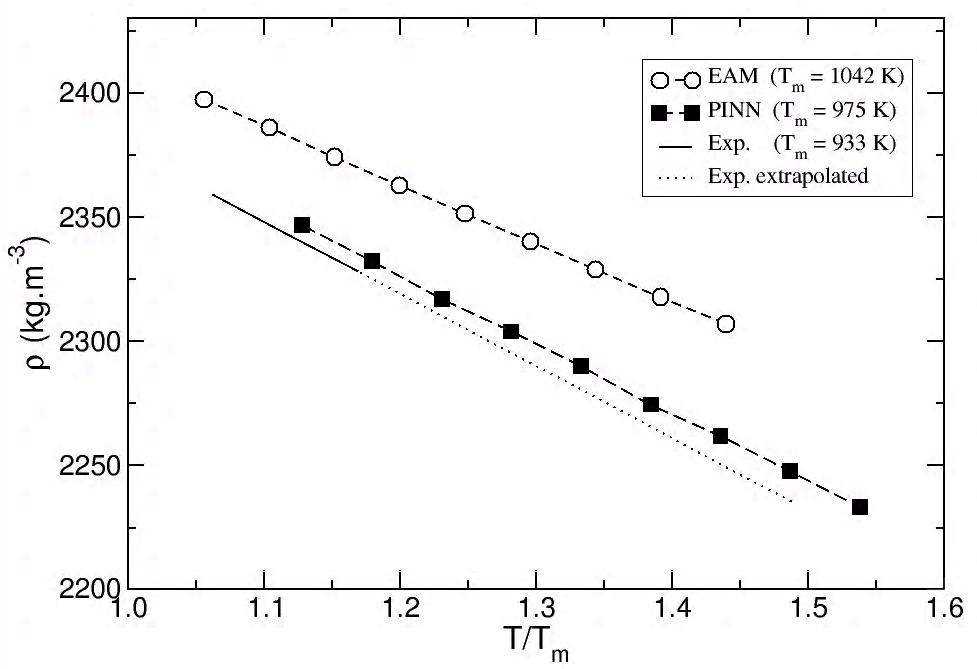}\caption{Density of liquid Al as a function of homologous temperature $T/T_{m}$
computed with the PINN and EAM \citep{Mishin99b} potentials in comparison
with experimental data \citep{Assael:2006aa}. \label{fig:density}}
\end{figure}

Figure \ref{fig:density} shows the temperature dependence of the
liquid density computed with the PINN and EAM \citep{Mishin99b} potentials
and measured experimentally \citep{Assael:2006aa}. To facilitate
comparison, the homologous temperature ($T/T_{m}$) is used as the
melting points predicted by the potentials are shifted relative to
the experiment. The PINN potential is clearly in better agreement
with experiment than EAM.

\begin{figure}[htpb]
\noindent \begin{centering}
\textbf{(a)} \includegraphics[width=0.6\textwidth]{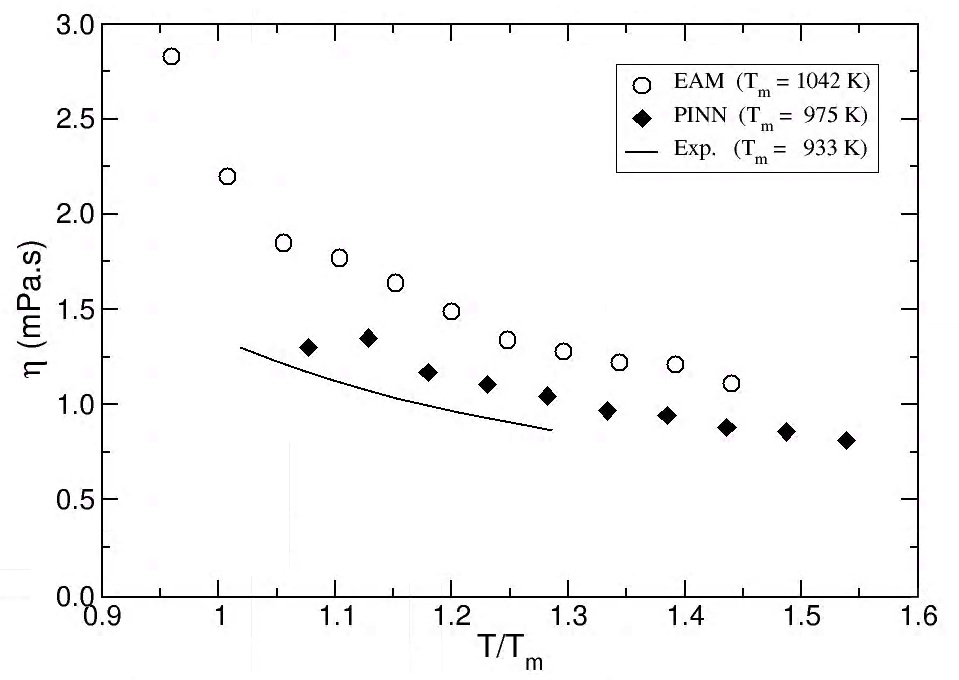}
\par\end{centering}
\bigskip{}

\bigskip{}

\noindent \centering{}\textbf{(b)} \includegraphics[width=0.6\textwidth]{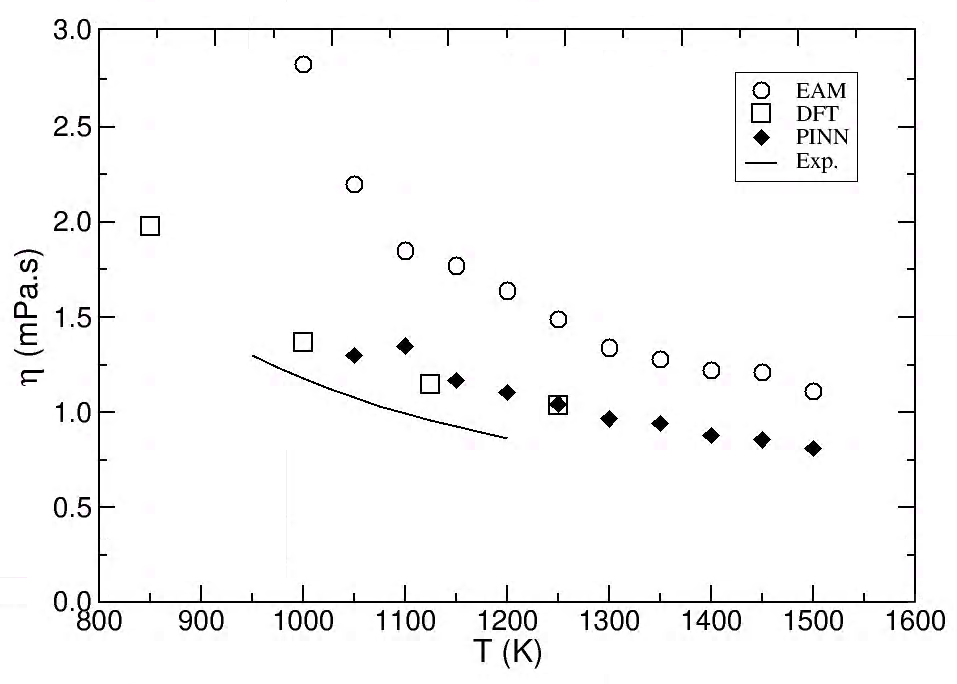}
\caption{Viscosity of liquid Al as a function of temperature computed with
the PINN and EAM \citep{Mishin99b} potentials in comparison with
(a) experimental data \citep{Assael:2006aa} using the homologous
temperature $T/T_{m}$, and (b) DFT calculations \citep{Jakse:2013aa}
using the actual temperature. \label{fig:viscosity}}
\end{figure}

For viscosity, the agreement with experiment \citep{Assael:2006aa}
is similarly good as illustrated in Fig.~\ref{fig:viscosity}a. Furthermore,
our results can be compared with DFT data reported by Jakse et al.~\citep{Jakse:2013aa}.
Their calculations were performed in both the local density approximation
(LDA) and the generalized gradient approximation (GGA). We choose
the GGA data for comparison because the DFT database \citep{Botu:2015aa,Botu:2015bb,PINN-1}
utilized for training and validation of the PINN potential was generated
in the GGA. We use the actual (not homologous) temperature because
the DFT melting temperature is unknown. Figure \ref{fig:viscosity}b
demonstrates that the PINN calculations are in excellent agreement
with the DFT data.

\begin{figure}[htpb]
\noindent \centering{}\includegraphics[width=0.7\textwidth]{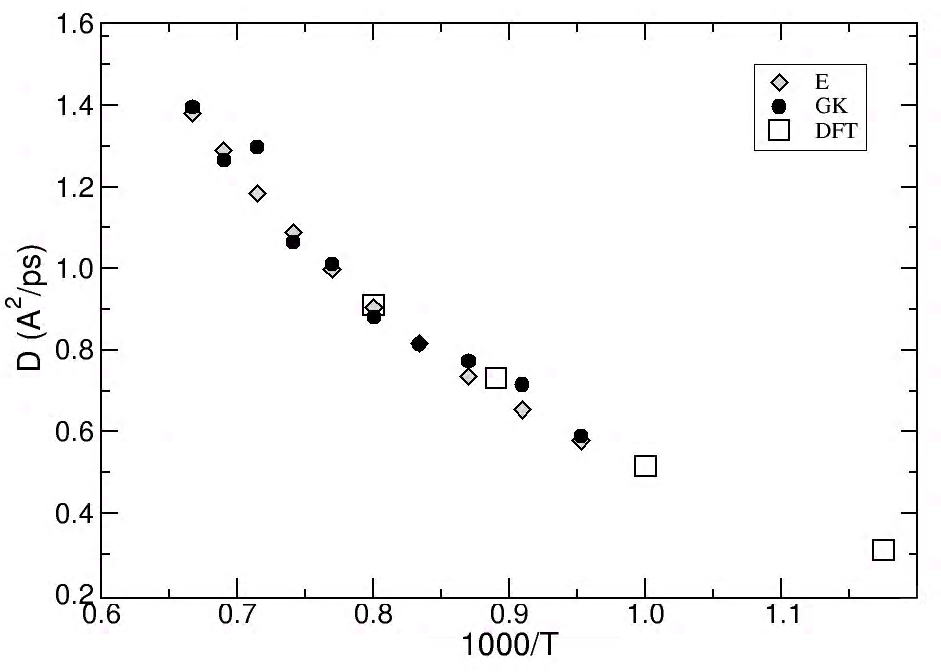}\caption{Arrhenius diagram of self-diffusion coefficients in liquid Al computed
with the PINN potential using the Green-Kubo (GK) and Einstein (E)
methods in comparison with DFT calculations \citep{Jakse:2013aa}.
\label{fig:diffusion}}
\end{figure}

Finally, the Arrhenius diagram of diffusion coefficients in liquid
Al is shown in Fig.~\ref{fig:diffusion}. Excellent agreement is
observed between the diffusivities obtained by the Green-Kubo and
Einstein methods. Equally good is the agreement between the PINN and
DFT calculations (again, using the GGA data) across the temperatures
covered by the simulations. Comparison with experiment has not been
attempted because the existing experimental data is not reliable enough
for a meaningful comparison. Accurate self-diffusion measurements
are made with stable or radioactive isotopes. Aluminum does not have
a suitable isotope for diffusion measurements. Hence the diffusivities
reported in the literature were obtained by indirect methods that
are less reliable.

\subsection{Grain boundary crack growth\label{subsec:GB-crack}}

To further demonstrate the possibility of conducting large-scale simulations
with PINN potentials, we performed simulations of a bicrystal containing
a crack growing on a grain boundary (GB) subject to an applied stress.
The system setup closely follows the one reported in a previous paper
\citep{Yamakov06}, where an EAM Al potential was used. Relying on
an already studied system helped us in establishing the correct loading
conditions ensuring a continuous crack growth after nucleation.

Crystallographic orientations of the two grains are $x:[\overline{7}\thinspace7\thinspace10]$,
$y:[5\thinspace\overline{5}\thinspace7]$, $z:[1\thinspace1\thinspace0]$
in the upper grain and $x:[\overline{7}\thinspace7\thinspace10]$,
$y:[\overline{5}\thinspace5\thinspace\overline{7}]$, $z:[1\thinspace1\thinspace0]$
in the lower grain. Thus, the two lattices are mirror images of each
other with respect to the GB plane $\left\{ 5\thinspace5\thinspace7\right\} $.
This GB is classified as $\Sigma99$ $[1\thinspace1\thinspace0]$
symmetrical tilt boundary with the misorientation angle of $89.42^{\circ}$
($\Sigma$ is the reciprocal density of coincident sites and $[1\thinspace1\thinspace0]$
is the tilt axis). The atomic structure of this boundary is known
from previous simulations and observations by atomic-resolution electron
microscopy \citep{Dahmen:1990aa}. The system thickness in the $z$-direction
is 10 $\left\{ 2\thinspace2\thinspace0\right\} $ crystallographic
planes ($\sim29$\,Å). This thickness is more than a factor of 4
larger than the cutoff radii of the PINN and EAM potentials tested
here, preventing interactions of atoms with their periodic images
and preserving the local three-dimensional physics. The system dimensions
in the $x$ and $y$ directions are 530\,Å and 497\,Å, respectively,
and the total number of atoms is 427,333.

Following thermal equilibration at the temperature of 100 K and zero
pressure, the system was loaded hydrostatically in tension to 4 GPa
and re-equilibrated at this stress. Once equilibrium was established
between the strain in the system and the applied external stress,
the system size in all three dimensions was fixed, creating a constant
strain condition. To save computer time, the equilibration steps were
first implemented with the EAM potential \citep{Mishin99b}. Transition
to the PINN potential was accomplished by an additional simulation
at constant temperature and strain for about 20 ps. After reaching
equilibrium with the PINN potential, a crack was nucleated by cutting
atomic interactions between atoms across the GB plane in 100 Å long
region. This length is larger than the Griffith length, $L_{G}\approx53$
Å, estimated for these loading conditions \citep{Yamakov06}. This
condition ensures that the crack will nucleate and grow, rather than
shrink and disappear.

\begin{figure}[htpb]
\noindent \begin{centering}
\includegraphics[angle=-90,width=0.5\textwidth]{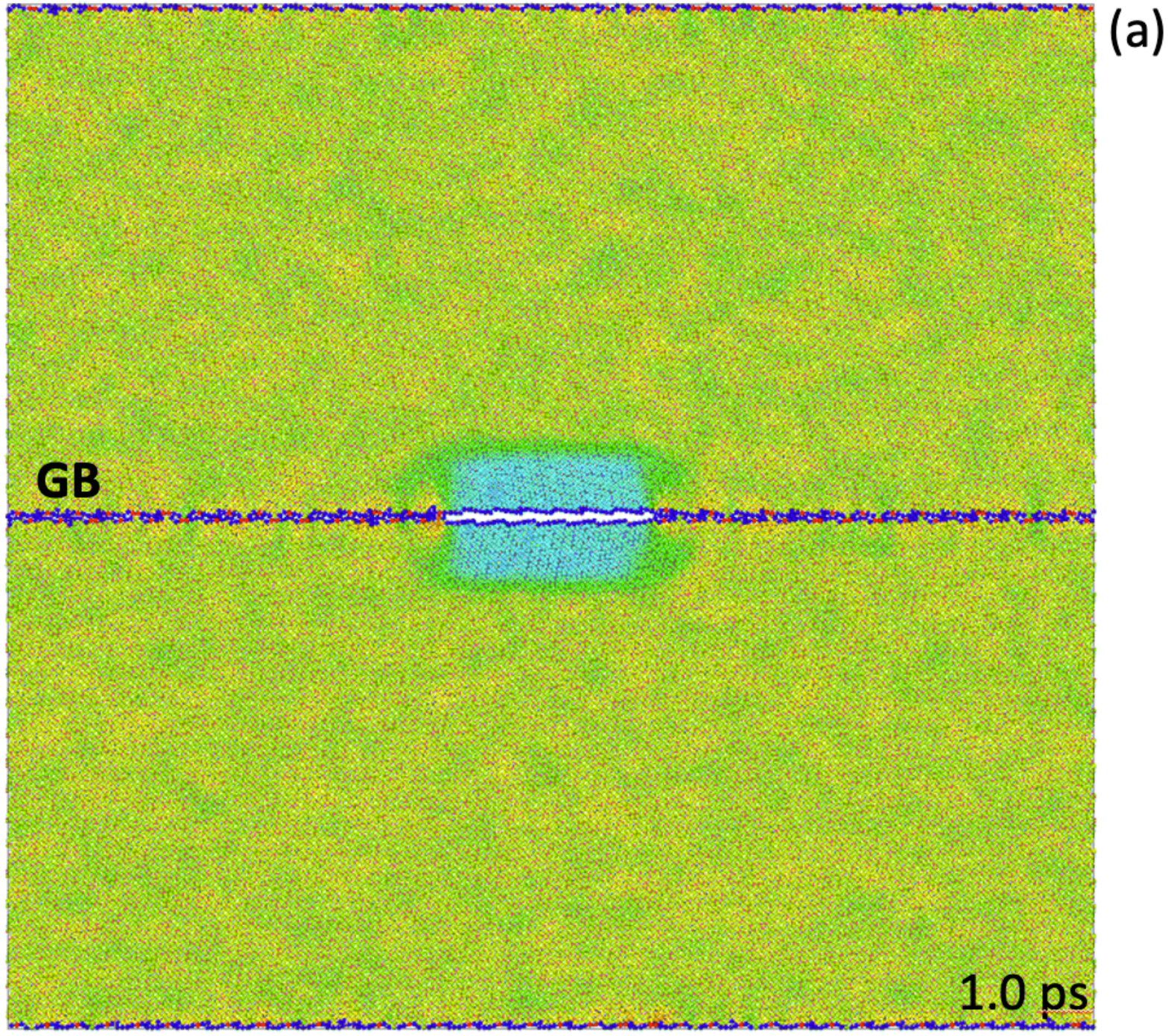}
\par\end{centering}
\bigskip{}

\noindent \begin{centering}
\includegraphics[angle=-90,width=0.5\textwidth]{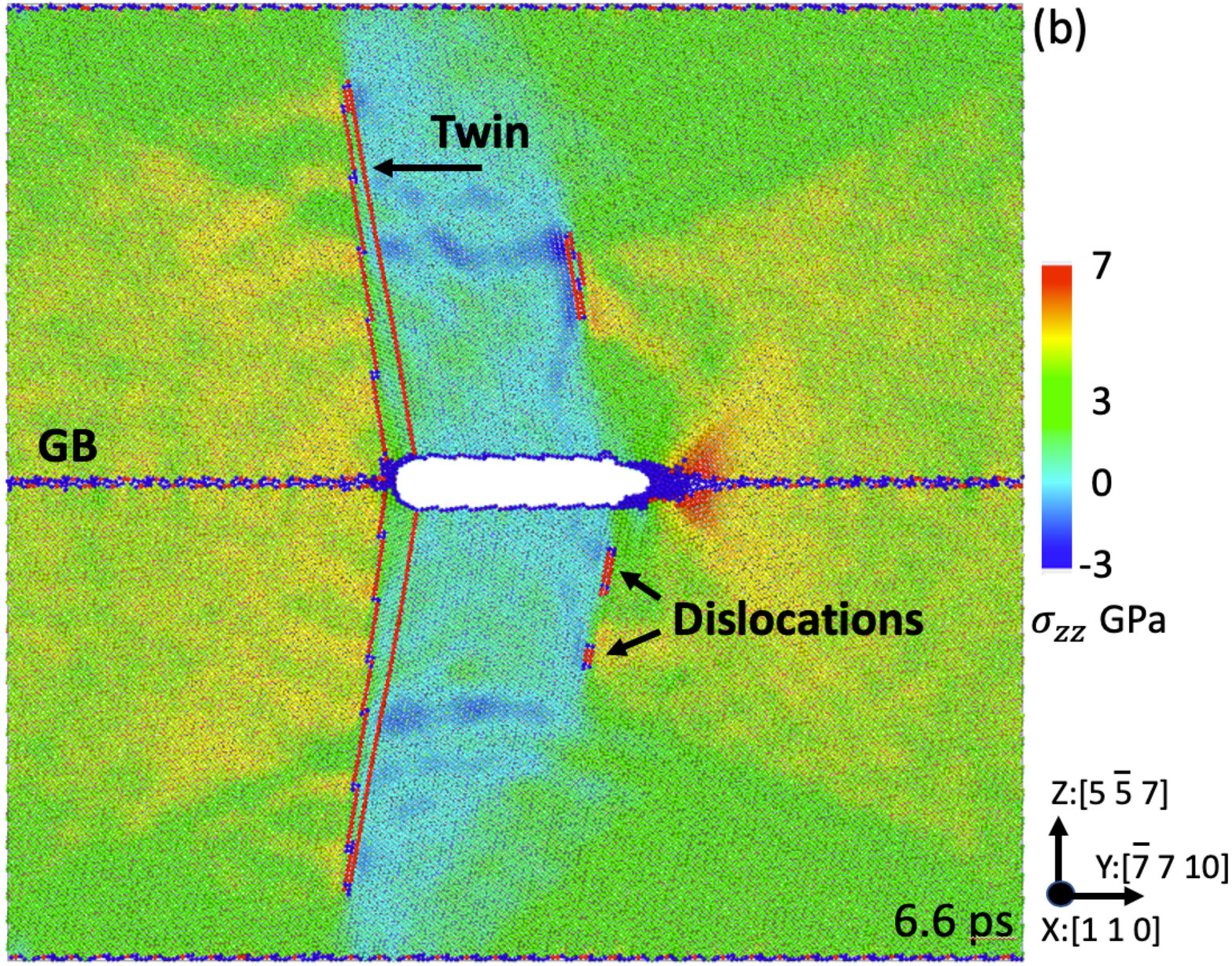}
\par\end{centering}
\bigskip{}

\noindent \begin{centering}
\includegraphics[angle=-90,width=0.5\textwidth]{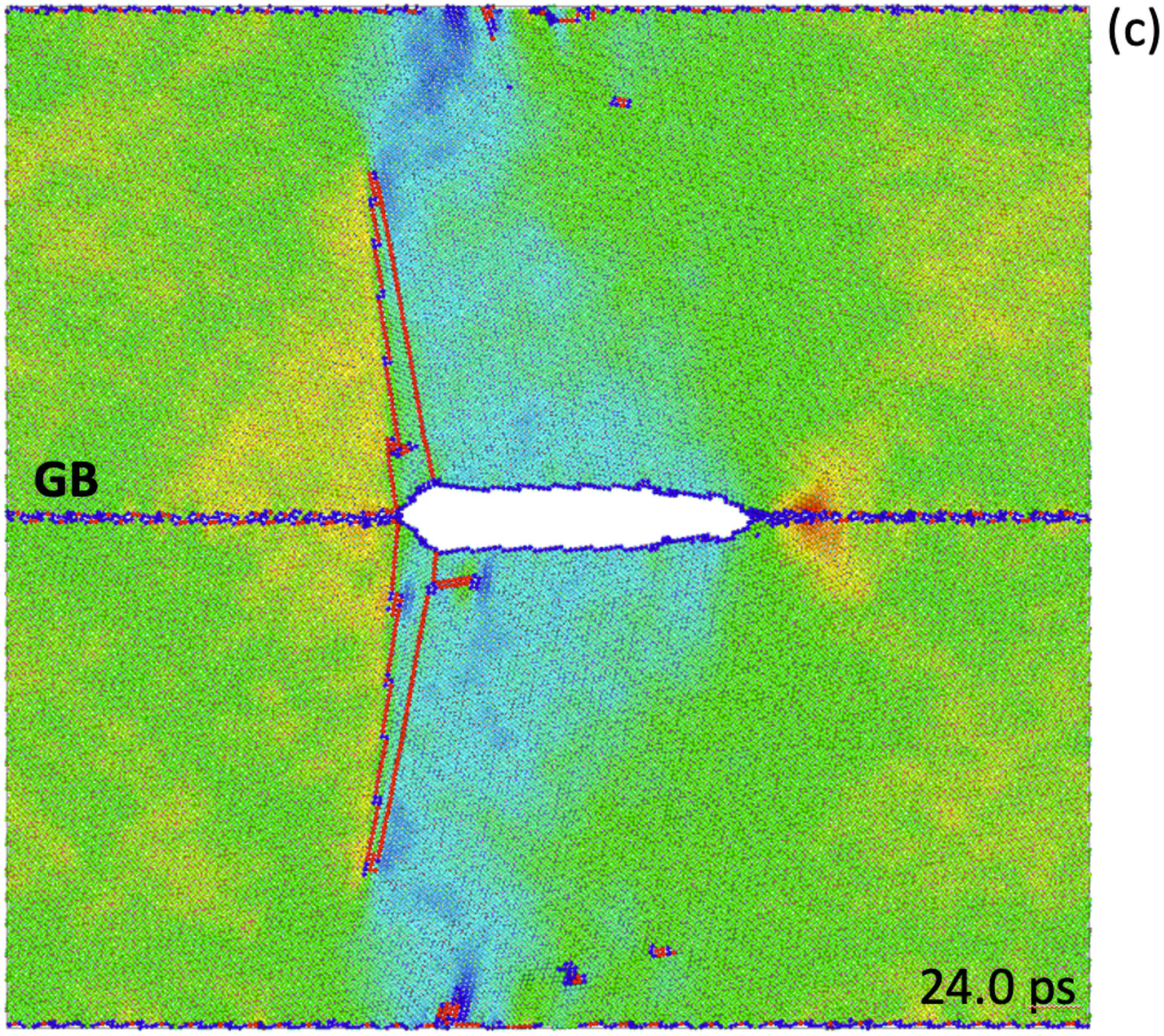}
\par\end{centering}
\noindent \centering{}\medskip{}
\caption{MD simulation of crack nucleation and growth on a $\Sigma99$ $[1\thinspace1\thinspace0]$
symmetrical tilt boundary in Al performed with the PINN Al potential.
(a) Early stage after crack nucleation; (b) Crack shape after 6.6
ps of growth; (c) End of the crack growth at 24 ps after nucleation.
Visualization of dislocations and twins is based on common neighbor
analysis superimposed on a tensile stress map using the opens source
code OVITO \citep{Stukowski2010a}. \label{fig:crack}}
\end{figure}

The snapshots in Figure \ref{fig:crack} represent the crack configurations
from the early stages of the simulation and during the growth for
24 ps of NVT MD time. The snapshots combine structural common neighbor
analysis (CNA) to identify the dislocations and twins, with the tensile
stress field given as a background. The simulation took approximately
14 h of CPU time on a CPU-GPU system described in the next Section.
The crack growth follows different mechanisms of propagation at the
two crack tips, depending on the inclination of (111) slip planes
with respect to the crack growth direction, such as the deformation
twinning on the left and dislocation emission on the right. The results
are fully consistent with the theoretical analysis \citep{Yamakov06}
predicting the different crack propagation mechanisms (dislocation
emission versus cleavage) based on the Rice criterion \citep{Rice1992}.
While the present simulations did not show notable differences from
the results in Ref.~\citep{Yamakov06}, they illustrate the capability
of the PINN potential to be used in simulations of the same scale
as with the traditional potentials such as EAM.

\section{Computational efficiency\label{sec:Computational-efficiency}}

The greatest advantage of ML potentials is their ability to accelerate
and upscale atomistic simulations relative to straight DFT calculations
while keeping a near-DFT level of accuracy in predicting the energy
and forces. A detailed comparison of computational performance of
ML potentials has recently been published \citep{Zuo:2020aa}. Like
other ML potentials, the PINN potentials scale linearly with the number
of atoms and are much faster than DFT calculations, but of course
slower than traditional potentials. Specific numbers depend not only
on the particular potential but also on the simulation software and
computer hardware. A few examples discussed below are only intended
to give a general idea about the computational efficiency of PINN.
These numbers may vary if a different simulation package and/or a
different computer architecture are used.

The training of the PINN Al potential reported here was performed
using an in-house code written in the C/C++ programming language and
parallelized with the Message Passing Interface (MPI). A typical training
run engaged 400 Central Processing Units (CPU) ((20 nodes)$\times$(20
cores each)) and took about an hour to complete 200 optimization iterations.
A complete optimization down to (2 to 3) meV per atom typically required
more than 4,000 iterations. As mentioned above, the optimization had
to be repeated multiple times to find an optimal combination of physical
properties and perform cross-validation.

The MD simulations were performed with a version of the PGMC code
\citep{ParaGrandMC,Purja-Pun:2015aa,Yamakov:2015aa}. The code is
parallelized by implementing a spatial decomposition that distributes
the system over a number of compute nodes connected through MPI. On
each node, an Open Multi-Processing (OpenMP) programming interface
was used to distribute the calculations over the available CPU cores.
When a Graphic Processing Unit (GPU) was available, the Open Accelerators
(OpenACC) programming interface was used to upload part of the calculations
on the GPU. In this case, the search for neighbors within the cutoff
range was performed with OpenMP taking advantage of all CPU cores,
while the energy and force calculations were uploaded to the GPU using
OpenACC. Accordingly, two computing configurations were used with
a similar performance: A CPU-only configuration, and a CPU-GPU configuration.
The latter consisted of a single node equipped with two dual socket
20 core Intel Gold 6148 Skylake CPU cores running at 2.40 GHz with
4 Nvidia V100 GPU cards (total: CPUs = 40, GPUs = 4). To utilize the
node architecture efficiently, MPI was used to spatially decompose
the system into 4 subdomains and represent the compute node as 4 MPI
nodes with 10 CPU cores and one GPU card each.

MD calculations of the melting point, interface tensions and the liquid
structure were conducted in the CPU-only mode using a single node
composed of 28 cores (1 MPI process with a total of 56 threads). A
typical MD simulation of 10,976 atoms took about 24 hrs to complete
40,000 MD steps (40 ps). In the liquid density, viscosity and diffusion
calculations, a 120 ps MD simulation took 47 to 72 CPU hrs (depending
on the machine load) using either the CPU-only configuration (8 MPI
nodes of 16 cores each) or the CPU-GPU configuration as already described.
Both configurations showed a similar computational performance. In
the GB crack simulation (427,333 atoms), the 24 ps MD simulation (12,000
MD steps) took about 14 hrs on the CPU-GPU system.

To evaluate the PINN efficiency relative to traditional potentials,
the EAM Al potential \citep{Mishin99b} was used as an example. With
the same PGMC code and the same computer hardware, the EAM potential
was found to be about a factor of 170 faster. Most of the overhead
time of PINN (about 65 \%) is spent on computing the local structural
descriptors. This step is common to all ML potentials, including the
purely mathematical NN potentials mentioned in Section \ref{sec:Introduction}.
The additional overhead due to the incorporation of the BOP potential
in PINN constitutes about 25 \% of the total time. The NN calculations
are the fastest taking less than 10 \% of the compute time. For comparison,
the integration of the equations of motion using the velocity Verlet
integrator take less than 0.1 \% of the compute time. Given the benefits
of the PINN approach discussed in the paper, we believe that this
modest overhead (about 25 \%) is worth its value.

\section{Conclusions\label{sec:Conclusions}}

The PINN model \citep{PINN-1} has been modified to accelerate the
potential training process and improve the transferability. Instead
of predicting the BOP parameters directly, the NN now predicts local
corrections to fixed parameters of a global BOP potential pre-trained
on the same DFT database. Such corrections (perturbations) play a
supporting role, whereas the physics-based global BOP potential takes
the lead in guiding the energy and forces. The on-the-fly adaptivity
through local corrections drastically improves the accuracy of the
potential, as illustrated by the BOP-PINN comparison in Fig.~\ref{fig:BOP-DFT}.
As long as the corrections remain relatively small, transferability
to unknown atomic environments must be robust. Of course, as with
any model, the PINN model eventually fails when the atomic configurations
arising during the simulations drift too far away from the training
domain and the predicted BOP parameters become unphysical. However,
the incorporation of physics through the BOP potential significantly
expands the range of validity of the potential in comparison with
purely mathematical ML potentials.

\begin{figure}[htpb]
\noindent \begin{centering}
\textbf{(a)} \includegraphics[scale=0.35]{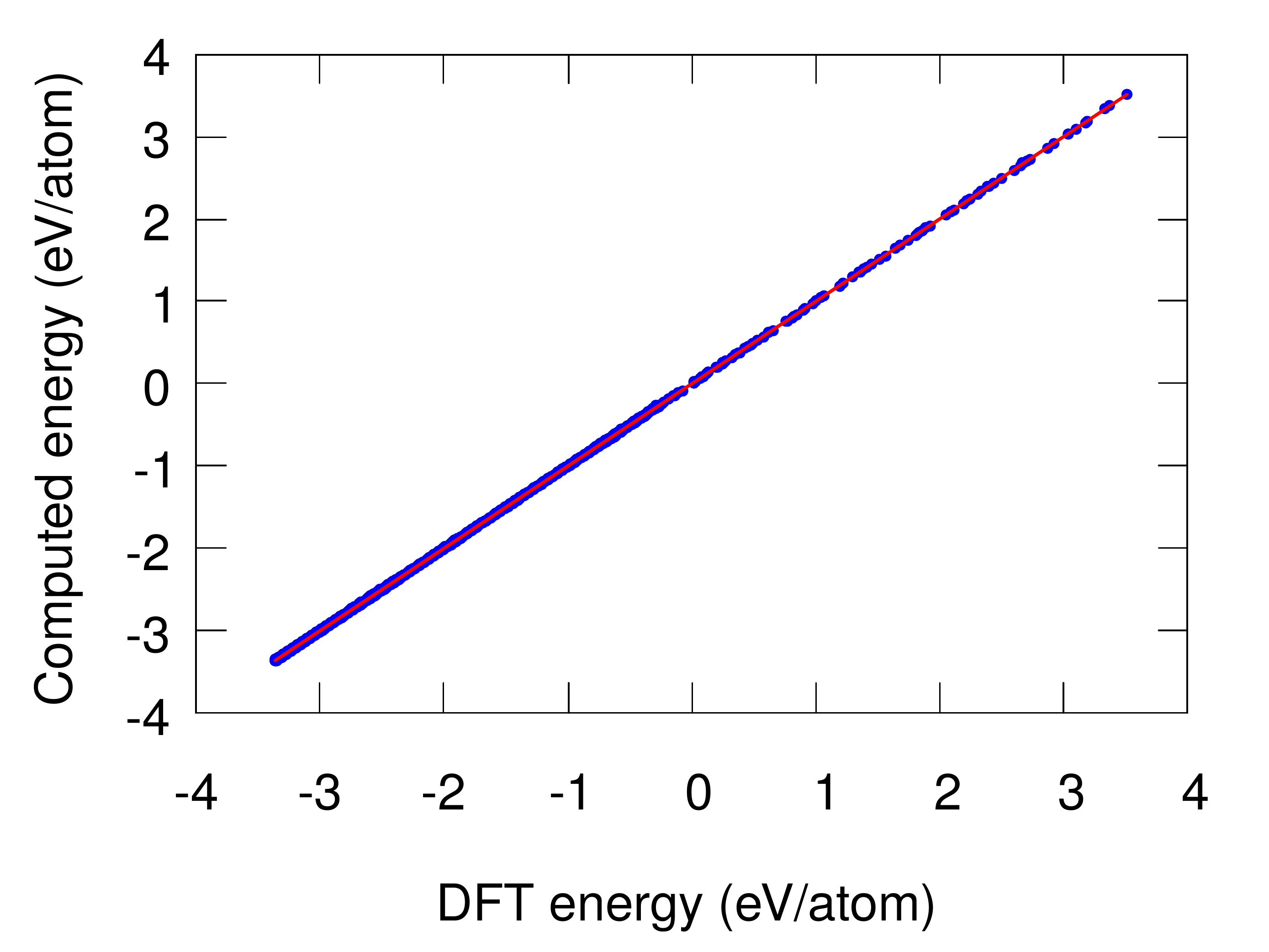}
\par\end{centering}
\bigskip{}

\bigskip{}

\noindent \begin{centering}
\textbf{(b)} \includegraphics[scale=0.35]{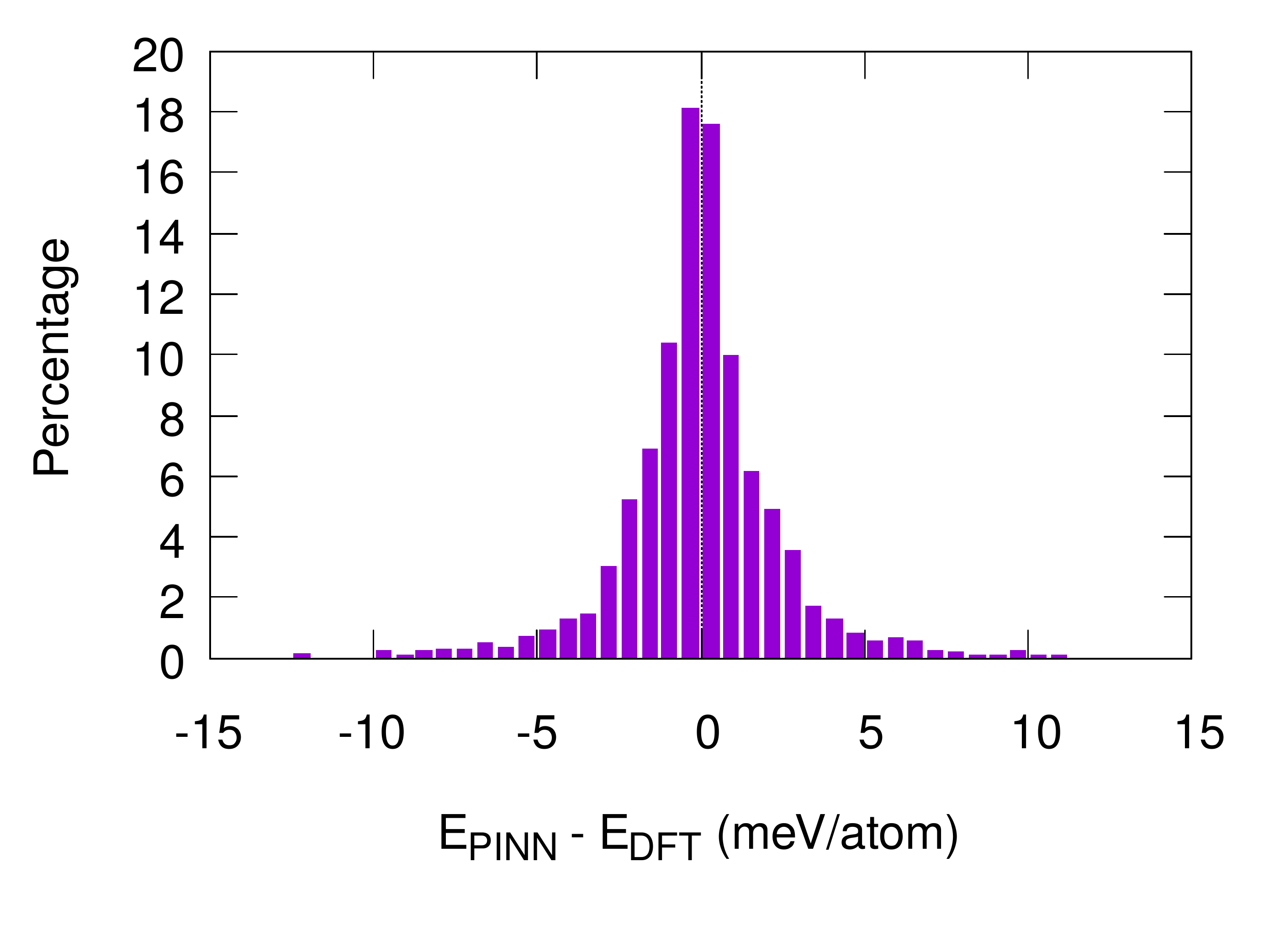}
\par\end{centering}
\noindent \centering{}\bigskip{}
\caption{(a) Energies computed with the PINN potential versus DFT energies
for the training database. The straight line represents perfect fit.
(b) Error distribution of the PINN potential.\label{fig:compVsDFT}}
\end{figure}

\begin{figure}[htpb]
\noindent \begin{centering}
\includegraphics[scale=0.4]{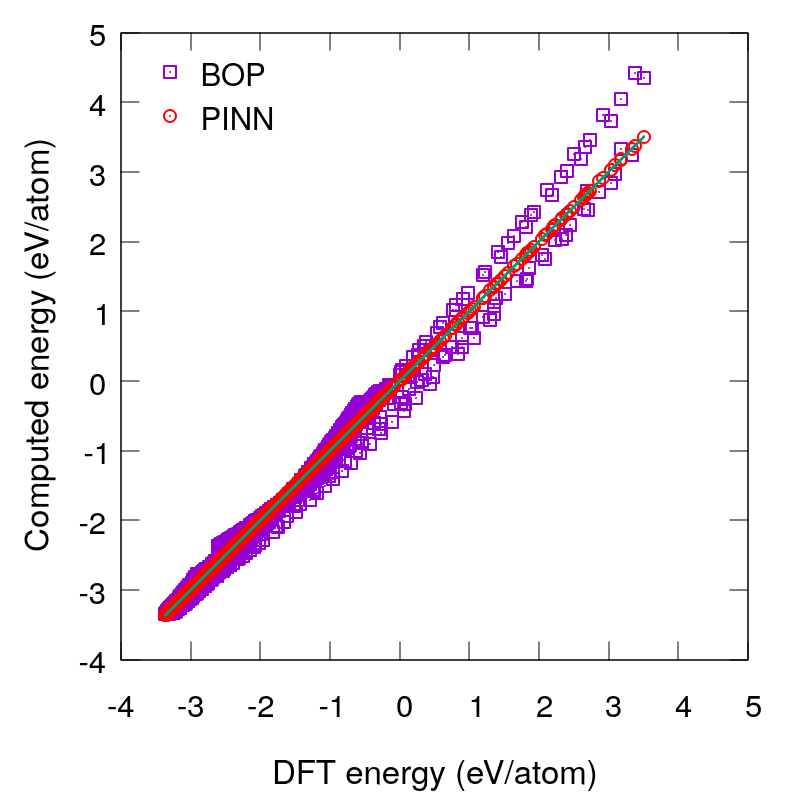}
\par\end{centering}
\noindent \centering{}\caption{Energies computed with the global BOP potential versus DFT calculations
for structures included in the training database. The PINN-DFT plot
from Fig.~\ref{fig:compVsDFT} is added for comparison.\label{fig:BOP-DFT}}
\end{figure}

As an application, a general-purpose Al potential has been constructed
following the modified PINN formalism. The potential accurately reproduces
the training DFT database (RMSE < 3 meV per atom) over a 7 eV per
atom wide energy range as shown in Fig.~\ref{fig:compVsDFT}. By
contrast to most of the existing ML potentials, the PINN potential
has been tested for a wide spectrum of physical properties. In fact,
it has been tested at least as thoroughly as traditional potentials
are normally tested prior to release to users. However, by contrast
to traditional potentials, the PINN Al potential demonstrates much
higher accuracy comparable to that of DFT calculations. The tests
have shown that the potential faithfully reproduces many properties
of Al obtained by DFT calculations (mostly collected from the literature).
When appropriate, comparison with experiment has been made and the
agreement was found to be reasonable. It should be noted that deviations
from experiment are partially accounted for by the fact that the potential
was trained on DFT data without any experimental input. DFT calculations
would not necessarily reproduce the experiment accurately either.
We include the comparison with experiment primarily to inform those
users who will be mainly interested in the ability of the potential
to reproduce or predict experimental data. This is often the case
in simulations geared towards practical applications.

The capability of the potential to perform large-scale simulations
has been demonstrated by computing the melting temperature of Al,
the structure and dynamic properties of liquid Al, the interface tensions
by the capillary fluctuation method, and the nucleation and growth
of a grain boundary crack. Some of these simulations involved tens
or hundreds of thousands of atoms and/or required MD runs for hundreds
of picoseconds. It should also be emphasized that these simulations
explored atomic environments that were significantly different from
those represented in the training database. As such, they mainly occurred
in the extrapolation regime.

Computational efficiency of ML potentials is an important factor that
drives their development and applications. ML potentials are orders
of magnitude faster than straight DFT calculations, but of course
slower than traditional potentials. Specifically, the PINN Al potential
developed here is estimated to be two order of magnitude slower than
a typical EAM potential. To maintain access to the same size of simulations,
more powerful computational resources and more efficient training
and simulation codes must be developed. Using the PGMC simulation
code as an example, it has been demonstrated that this goal can be
achieved by a proper combination of parallel programming interfaces
highly optimized for the available computer architectures.

The current work includes the incorporation of PINN potentials into
other large-scale simulation packages, such as the Large-scale Atomic/Molecular
Massively Parallel Simulator (LAMMPS) \citep{Plimpton95}, construction
of PINN potentials for other metallic and nonmetallic materials, and
the development of a multi-component version of PINN. In the latter
case, the size $K$ of the feature vector $(G_{i}^{1},G_{i}^{2},...,G_{i}^{K})$
grows as the number of chemical components squared, which is a common
property of all ML potentials. The number of BOP parameters also scales
up in the same proportion, resulting in a drastic increase in the
number of weights and biases in the network. The size of the training
database must also be much larger to properly represent different
chemical compositions of the system. Even the development of a binary
PINN potential is a demanding task, but can still be accomplished
given sufficient computational resources. Several ML potentials for
binary and ternary systems have already been been reported in the
literature \citep{Artrith:2011ab,Artrith:2011aa,Sosso2012,Artrith:2016aa,Artrith:2017aa,Hajinazar:2017aa,Kobayashi:2017aa,Artrith:2018aa,Li:2018aa,Hajinazar:2019aa,Gubaev:2019aa,Zhang:2019ab}.
\begin{acknowledgments}
We are grateful to R.~Batra and R.~Ramprasad for making the DFT
Al database available for the development of PINN potentials. G.~P.~P~and
Y.~M.~were supported by the Office of Naval Research under Awards
No.~N00014-18-1-2612. V.~Y.~was sponsored through a NASA cooperative
agreement NNL09AA00A with the National Institute of Aerospace. The
work of V.~Y.~and E.~H.~G.~was supported by NASA's Transformational
Tools and Technologies project. J.~H.~was supported by an NRC Research
Associateship award at the National Institute of Standards and Technology
(NIST).
\end{acknowledgments}

\newpage\clearpage{}


\newpage\clearpage{}

\begin{table}[thpb]
\caption{Aluminum properties predicted by the PINN potential in comparison
with experimental data and DFT calculations. $E_{0}$ - equilibrium
cohesive energy, $a_{0}$ - equilibrium lattice parameter, $B$ -
bulk modulus, $c_{ij}$ - elastic constants, $\gamma_{s}$ - surface
energy, $E_{v}^{f}$ - vacancy formation energy, $E_{v}^{m}$ - vacancy
migration barrier, $E_{I}^{f}$ - interstitial formation energy for
the tetrahedral ($T_{d}$) and octahedral ($O_{h}$) positions and
split dumbbell configurations with different orientations, $\gamma_{\textrm{SF}}$
- intrinsic stacking fault energy, $\gamma_{\textrm{us}}$ - unstable
stacking fault energy. All defect energies are statically relaxed
unless otherwise indicated. \label{table:al_prop}}

\bigskip{}
\begin{tabular}{llccc}
\hline 
Property &  & DFT &  & PINN\tabularnewline
\hline 
$E_{0}$ (eV per atom) &  & 3.7480$^{a}$ &  & 3.3604\tabularnewline
$a_{0}$ (Å) &  & 4.039$^{a,d}$; 3.9725--4.0676$^{c}$ &  & 4.0399\tabularnewline
$B$ (GPa) &  & 83$^{a}$; 81$^{f}$ &  & 81\tabularnewline
$c_{11}$ (GPa) &  & 104$^{a}$; 103--106$^{d}$ &  & 112\tabularnewline
$c_{12}$ (GPa) &  & 73$^{a}$; 57--66$^{d}$ &  & 65\tabularnewline
$c_{44}$ (GPa) &  & 32$^{a}$; 28--33$^{d}$ &  & 28\tabularnewline
$\gamma_{s}$(100) (J\,m$^{-2}$) &  & 0.92$^{b}$ &  & 0.904\tabularnewline
$\gamma_{s}$(110) (J\,m$^{-2}$) &  & 0.98$^{b}$ &  & 0.954\tabularnewline
$\gamma_{s}$(111) (J\,m$^{-2}$) &  & 0.80$^{b}$ &  & 0.804\tabularnewline
$E_{v}^{f}$ (eV) &  & 0.6646--1.3458$^{c}$; 0.7$^{e}$ &  & 0.703\tabularnewline
$E_{v}^{f}$ (eV) unrelaxed &  & 0.78$^{e}$ &  & 0.76\tabularnewline
$E_{v}^{m}$ (eV) &  & 0.3041--0.6251$^{c}$; &  & 0.628\tabularnewline
$E_{I}^{f}$ ($T_{d}$) (eV) &  & 2.2001--3.2941$^{c}$ &  & 2.760\tabularnewline
$E_{I}^{f}$ ($O_{h}$) (eV) &  & 2.5313--2.9485$^{c}$ &  & 2.739\tabularnewline
$E_{I}^{f}$ $\langle$100$\rangle$ (eV) &  & 2.2953--2.6073$^{c}$ &  & 2.517\tabularnewline
$E_{I}^{f}$ $\langle$110$\rangle$ (eV) &  & 2.5432--2.9809$^{c}$ &  & 2.843\tabularnewline
$E_{I}^{f}$ $\langle$111$\rangle$ (eV) &  & 2.6793--3.1821$^{c}$ &  & 2.775\tabularnewline
$\gamma_{\textrm{SF}}$ (mJ\,m$^{-2}$) &  & 134$^{i}$; 145.67$^{g}$; 158$^{h}$ &  & 134\tabularnewline
$\gamma_{\textrm{us}}$ (mJ\,m$^{-2}$) &  & 162$^{j}$; 175$^{h}$ &  & 150\tabularnewline
\hline 
\hline 
\multicolumn{3}{l}{$^{a}$ Ref.\,\citep{Jong:2015fk}; $^{b}$ Ref.\,\citep{Tran:2016qq};
$^{c}$ Ref.\,\citep{Qiu:2017ve}; $^{d}$ Ref.\,\citep{Zhuang:2016}
$^{e}$ Ref.\,\citep{Iyer:2014};} & \tabularnewline
\multicolumn{3}{l}{$^{f}$ Ref.\,\citep{Sjostrom:2016}; $^{g}$ Ref.\,\citep{Devlin:1974qp};$^{h}$
Ref.\,\citep{OgataLY02}; $^{i}$ Ref.\,\citep{Jahnatek:2009aa};
$^{j}$ Ref.\,\citep{Kibey:2007aa}} & \tabularnewline
\end{tabular}
\end{table}

\begin{table}[hptb]
\caption{Surface tension of liquid Al predicted by the PINN and EAM \citep{Mishin99b}
potentials at the respective melting temperatures. Experimental results
measured on microscopic droplets are included for comparison.\label{table:surface_tension}}

\bigskip{}
\begin{tabular}{|l|c|c|c|c|}
\hline 
Method & System size & Number of atoms & \multicolumn{2}{c|}{$\gamma$ (J\,m$^{-2}$)}\tabularnewline
\cline{4-5} \cline{5-5} 
 &  &  & Capillary waves & Laplace pressure\tabularnewline
\hline 
\hline 
PINN & 622\,Å$\times$29\,Å$\times$196\,Å & 186,000 & 0.610 & 0.613\tabularnewline
\hline 
EAM & 619\,Å$\times$29\,Å$\times$359\,Å & 360,000 & 0.717 & 0.738\tabularnewline
\hline 
Experiment & $\approx$mm &  & \multicolumn{2}{c|}{0.828$^{a}$; 0.87$^{b}$}\tabularnewline
\hline 
\multicolumn{1}{l}{$^{a}$\,Ref.~\citep{Poirier:1987aa}} & \multicolumn{1}{c}{} & \multicolumn{1}{c}{} & \multicolumn{1}{c}{} & \multicolumn{1}{c}{}\tabularnewline
\multicolumn{1}{l}{$^{b}$\,Ref.~\citep{Egry:2008aa}} & \multicolumn{1}{c}{} & \multicolumn{1}{c}{} & \multicolumn{1}{c}{} & \multicolumn{1}{c}{}\tabularnewline
\end{tabular}
\end{table}

\begin{table}[hptb]
\caption{The stiffness $(\gamma+\gamma^{\prime\prime})$ of the solid-liquid
interface in Al computed with the PINN and EAM potentials at the respective
melting temperatures. Experimental data is included for comparison.\label{table:surface_tension-1}}

\bigskip{}
\begin{tabular}{|l|c|c|c|}
\hline 
Method & System size & Number of atoms & \multicolumn{1}{c|}{$(\gamma+\gamma^{\prime\prime})$ (mJ\,m$^{-2}$)}\tabularnewline
\hline 
PINN & 622\,Å$\times$29\,Å$\times$364\,Å & 360,000 & 95\tabularnewline
\hline 
EAM \citep{Mishin99b} & 619\,Å$\times$29\,Å$\times$359\,Å & 360,000 & 99\tabularnewline
\hline 
Other calculations &  &  & \multicolumn{1}{c|}{110$^{a}$; 135.2$^{b}$}\tabularnewline
\hline 
Experiment &  &  & $158\pm30$$^{c}$; 131-153$^{d}$\tabularnewline
\hline 
\multicolumn{2}{l}{$^{a}$\,Ref.~\citep{Morris02a} (EAM)} & \multicolumn{1}{c}{} & \multicolumn{1}{c}{}\tabularnewline
\multicolumn{2}{l}{$^{b}$\,Ref.~\citep{Asadi2015} (MEAM)} & \multicolumn{1}{c}{} & \multicolumn{1}{c}{}\tabularnewline
\multicolumn{2}{l}{$^{c}$\,Ref.~\citep{Jiang08} (Dihedral angle)} & \multicolumn{1}{c}{} & \multicolumn{1}{c}{}\tabularnewline
\multicolumn{3}{l}{$^{d}$\,Ref.~\citep{Jiang08} (Melting point depression)} & \multicolumn{1}{c}{}\tabularnewline
\end{tabular}
\end{table}

\newpage\clearpage{}

\end{document}